\documentclass[preprint,11pt]{aastex}

\shorttitle{VLA Protostars}
\shortauthors{Tobin et al.}

\newcommand{\nthp}{\mbox{N$_2$H$^+$}}

\newcommand{\nht}{\mbox{NH$_3$}}
\newcommand{\hcop}{\mbox{HCO$^+$}}

\newcommand{\kmspc}{\mbox{km s$^{-1}$ pc$^{-1}$ }}
\newcommand{\kms}{\mbox{km s$^{-1}$}}

\begin{document}

\title{VLA and CARMA Observations of Protostars in the Cepheus Clouds: Sub-arcsecond Proto-Binaries Formed via Disk Fragmentation }
\author{John J. Tobin\altaffilmark{1,}\altaffilmark{9}, Claire J. Chandler\altaffilmark{2} , David J. Wilner\altaffilmark{3},
Leslie W. Looney\altaffilmark{1,4},  Laurent Loinard\altaffilmark{5}, Hsin-Fang Chiang\altaffilmark{6}, 
Lee Hartmann\altaffilmark{7}, Nuria Calvet\altaffilmark{7}, Paola D'Alessio\altaffilmark{5}, Tyler L. Bourke\altaffilmark{3}, Woojin Kwon\altaffilmark{8}}

\altaffiltext{1}{National Radio Astronomy Observatory, Charlottesville, VA 22903; jtobin@nrao.edu}
\altaffiltext{2}{National Radio Astronomy Observatory, Socorro, NM}
\altaffiltext{3}{Harvard-Smithsonian Center for Astrophysics, Cambridge, MA 02138}
\altaffiltext{4}{Department of Astronomy, University of Illinois, Urbana, IL 61801 }
\altaffiltext{5}{Centro de Radioastronom{\'\i}a y Astrof{\'\i}sica, UNAM, Apartado Postal 3-72 (Xangari), 58089 Morelia, Michoac\'an, M\'exico}
\altaffiltext{6}{Institute for Astronomy \& NASA Astrobiology Institute, University of Hawaii at Manoa, Hilo, HI 96720}
\altaffiltext{7}{Department of Astronomy, University of Michigan, Ann Arbor, MI 48109}
\altaffiltext{8}{SRON Netherlands Institute for Space Research, Landleven 12, 9747 AD Groningen, The Netherlands}
\altaffiltext{9}{Hubble Fellow}

\begin{abstract}
We present observations of three Class 0/I protostars (L1157-mm, CB230 IRS1, and L1165-SMM1) using the  Karl G. Jansky Very Large Array (VLA) and observations of two (L1165-SMM1  and CB230 IRS1) with the Combined Array for Research in Millimeter-wave Astronomy  (CARMA). The VLA observations were taken at wavelengths of $\lambda = 7.3$ mm, 1.4 cm, 3.3 cm, 4.0 cm, and 6.5 cm with a best resolution of $\sim$0\farcs06 (18 AU) at 7.3 mm. The L1165-SMM1 CARMA observations were taken at $\lambda = 1.3$ mm with a best resolution of $\sim0\farcs3$ (100 AU), and the CB230 IRS1 observations were taken at $\lambda = 3.4$ mm with a best resolution of $\sim$3\arcsec\ (900 AU).  We find that L1165-SMM1 and CB230 IRS1 have probable binary companions at separations of $\sim$0\farcs3 (100 AU) from detections of secondary peaks at multiple wavelengths.  The position angles of these companions are nearly orthogonal to the direction of the observed bipolar outflows, consistent with the expected protostellar disk orientations. We suggest that these companions may have formed from disk fragmentation; turbulent fragmentation would not preferentially arrange the binary companions to be orthogonal to the outflow direction.  For L1165-SMM1, both the 7.3 mm and 1.3 mm emission  show evidence of a large (R $>$ 100 AU) disk. For the L1165-SMM1 primary protostar and the CB230 IRS1 secondary protostar, the 7.3 mm emission is resolved into structures consistent with $\sim20$ AU radius disks. For the other protostars, including L1157-mm, the emission is unresolved, suggesting disks with radii $< 20$ AU.

\end{abstract}


\section{Introduction}

Binary or multiple systems comprise at least half of all solar-type stars and the physical separation
of companions ranges from $\sim$ 10000 AU to $<$ $\sim$0.01 AU. The formation mechanism
for multiple systems is still debated \citep[e.g.][]{tohline2002}, but seems to occur during
the protostellar phase when there is still a significant mass reservoir available. Thus,
in order to test binary formation mechanisms, observations of very young sources that
have not undergone significant evolution are the most ideal. The earliest
recognized phase of protostellar evolution is the Class 0 phase \citep{andre1993};
 Class 0 sources are characterized by a large, dense envelope
surrounding the protostar(s). The Class I phase follows the Class 0 phase, where the
protostar is still embedded in its envelope, but the envelope has become less massive due to a
combination of infall and/or outflow dissipation \citep[e.g.][]{arce2006}.

Rotationally-supported disks\footnote{Note that the 
term ``disk'' used in the text implicitly refers to a rotationally-supported or Keplerian disk,
whether or not it has been kinematically verified.} were once thought to form as a simple consequence of angular momentum
conservation during the collapse of the protostellar envelope \citep[e.g.][]{ulrich1976,cassen1981,tsc1984}.
The angular momentum causes the infalling material to subsequently fall onto the disk and 
the disk can grow as higher angular momentum material falls in and becomes rotationally supported at
progressively larger radii \citep{yorke1999}. The accretion to the protostar
is then mediated by the circumstellar disk. Angular momentum is also thought to 
contribute to the formation of multiple systems by causing infalling material to form a 
rotationally flattened region in the inner envelope, significantly larger than a disk, where density perturbations would 
cause multiple collapse centers \citep{bb1993,boss1995,tohline2002}.

This simple picture of disk and binary formation breaks down in collapse models 
that consider magnetic braking \citep{mouschovias1979,basu1994}. Ideal magneto-hydrodynamic (MHD)
simulations and analytic models showed that even when gravity is dominant over the magnetic field in the starless phase,
the magnetic braking can slow the rotation during collapse phase and prevent the formation of
rotationally supported disks \citep{allen2003,galli2006,mellon2008,mellon2009}. The slowed rotation
on $\sim$1000 AU scales would also prevent binary formation from rotation-induced envelope 
fragmentation \citep[i.e.][]{bb1993,boss1995}. Moreover, given that disk formation is suppressed, the fragmentation
of these disks to form close binaries is also more difficult \citep{hennebelleteyssier2008} than in
the non-magnetic case \citep[e.g.][]{vorobyov2010, stamatellos2009}.
Later models considering dissipative, non-ideal MHD effects, do enable the formation of initially
very small disks \citep{dapp2010} and misalignment of 
magnetic fields with respect to the rotation axis of a system may also 
enable disk formation in the absence of strong dissipative effects \citep{joos2012, li2013}. 

Observationally, there had been little direct evidence for resolved disks 
toward Class 0 sources \citep{chiang2008,maury2010,chiang2012}
until a rotationally-supported disk, 125 AU in radius was 
discovered around the Class 0 protostar L1527 \citep{tobin2012}.
$^{13}$CO ($J=2\rightarrow1$) emission confirmed rotational support and enabled a measurement of
protostellar mass ($\sim$0.2 $M_{\sun}$). There is also the recent detection of a possible 
Keplerian disk around the prototype Class 0 system VLA1623 \citep{murillo2013}. 
Only a few disks have been resolved at millimeter wavelengths around Class I sources 
\citep[e.g.][]{rodriguez1998, launhardt2001c, brinch2007,lommen2008,jorgensen2009,takakuwa2012,hara2013},
but the results do point to large disks being prevalent in the Class I phase \citep{eisner2012}.

The multiplicity of protostars is better characterized than their
disk properties thus far. \citet{connelley2008} carried out a multiplicity survey of 
Class I protostars in the near-infrared (1.6 \micron, 2.15 \micron, and 3.7 \micron)
with spatial resolutions as fine as $\sim$ 70 AU (0\farcs33). They found a rather flat separation
distribution between 100 AU and 1000 AU, with increasing multiplicity on 3000 AU scales.
At separations $<$ 500 AU, an expected scale for disk formation, that study found 35 multiple protostars.
However, the infrared observations are biased against protostars that are deeply embedded 
in their natal envelopes, as most Class 0 and Class 0/I protostars are. 
Only interferometric observations in the millimeter
and centimeter can characterize multiplicity in the early stages of protostellar evolution.
Work by \citet{looney2000} showed that multiple systems are prevalent on 
spatial scales of $\ga$1000 AU in the Class 0 and I 
phases. \citet{chen2013} recently found that Class 0 protostars
might have a higher multiplicity fraction than Class I protostars.
There have only been a few studies with high enough resolution
to probe multiplicity on scales $<$ 500 AU in the millimeter; \citet{maury2010} had suggested
a lack of Class 0 binary systems between 150 AU and 550 AU based on their data combined with
\citet{looney2000}, arguing that fragmentation and disk formation were suppressed by
magnetic fields. Another study by \citet{enoch2011} in the 
Serpens star forming region only detected 1 source out of 9 with evidence 
of multiplicity on scales less than 415 AU. However, \citet{chen2013} found
evidence for 7 systems to be multiple with separations $<$ 500 AU and \citet{reipurth2002,reipurth2004} observed 21 systems
and found 7 to be multiple systems with 4 separated by $<$ 500 AU.

The results on scales $<$ 150 AU are quite sparse; there are currently only 5 Class 0 or Class I 
systems known to be multiples or at least candidate multiples
 with such close separations from millimeter and centimeter observations: 
L1551 NE \citep{reipurth2002}, L1551 IRS5 \citep{looney1997, rodriguez1998},
 IRAS 16293-2422A \citep{pech2010}, HH211 \citep{lee2009}, VLA1623 \citep{murillo2013}.
\citet{connelley2008} found 13 multiple Class I protostars with binary separations $<$ 150 AU
in the infrared. Thus far, it is unclear if the small
numbers at radio wavelengths reflect a true paucity or simply lack of observations with high enough resolution 
(or high resolution observations with low sensitivity). A paucity of binaries on scales $<$ 150 AU could indicate that binaries
typically form with larger separations ($\sim$few $\times$ 1000 AU) and migrate 
to smaller radii. \citet{offner2010} proposed such a scenario where turbulent
fragmentation caused binary systems to form and they subsequently migrated.

The conflicting and sparse results on binaries and disks in the early phases of protostellar
evolution can be attributed to small sample sizes and low spatial resolution.
The maximum angular resolution of sub/millimeter interferometers, 
has been previously limited to $\sim$0\farcs2, corresponding to a 
spatial resolution of $\sim$45 AU at the 230 pc distance
of the Perseus molecular cloud; however, few protostars have been 
observed at these highest resolutions needed to characterize
close binary protostars and disks. Spatial resolutions better than 50 AU are 
necessary in order to determine when (what evolutionary stage) and where
binary protostars form, and by what mechanism; 50 AU is roughly the peak
of the binary separation distribution for solar-type field stars \citep{raghavan2010}. It is difficult to determine
the fragmentation mechanism on an individual basis; however, the characteristics of the systems may 
indicate that one mechanism is more likely than another. Moreover, with the small sample of known
proto-binaries separated by $<$ 150 AU, even small additions are important.

The NRAO Karl G. Jansky Very Large Array (VLA) opens a new window 
for the study of protostars with its order of magnitude increase in the sensitivity 
in the 7 mm band and currently unprecedented resolution of $\sim$0\farcs06. We have observed
three Class 0/I protostars (L1157-mm, CB230, and L1165-SMM1) in the Cepheus Flare region (d $\sim$ 300 pc) 
\citep[e.g.][]{kirk2009}, attempting to characterize their disk properties and multiplicity.
Complementary data are taken with the Combined Array for Research in 
Millimeter-wave Astronomy (CARMA) for L1165-SMM1 and CB230 IRS1, with corresponding 
CARMA data taken from the literature for L1157-mm \citep{chiang2012}. Since all these targets are at declinations 
$>$ 59\degr, they cannot be observed by the nearly complete Atacama Large 
Millimeter/submillimeter Array (ALMA) and these VLA observations 
provide the most detailed view of their small-scale structure for the 
foreseeable future. Despite the small sample of protostars, the high-fidelity dataset enables us to 
shed new light on the properties of disks and multiplicity in young protostellar objects. 

The paper is organized as follows: the sample, observations, and data reduction
are detailed in Section 2, the basic observational results are given in Section 3, the 
results are discussed in Section 4, and we present our conclusions in Section 5.

\section{The Sample, Observations, Data Reduction, and Analysis}

We observed the protostars CB230 IRS1, L1165-SMM1, and L1157-mm with the VLA in A, B, and C
configurations at $\lambda$ = 7.3 mm, 1.4 cm, 3.3 cm, 4.0 cm, and 6.5 cm; the source properties and coordinates are given in Table 1.
We also obtained observations from CARMA for L1165-SMM1 at 1.3 mm in B and C configurations and CB230 IRS1 at 3.4 mm in C configuration.

\subsection{Sources}
The three selected protostars are all isolated and located in the Cepheus Flare region.
L1157-mm is found within one of several denser regions within a larger ring-like molecular cloud \citep{kirk2009} and
L1165-SMM1 and CB230 IRS1 reside within discrete dark clouds.
Despite the small sample, the sources span a range of luminosities and bolometric temperatures. 
Class 0 and Class I protostars are typically distinguished using the
ratio of submillimeter luminosity ($L_{submm}$) to bolometric luminosity 
($L_{bol}$) and/or bolometric temperature ($T_{bol}$).
The bolometric temperature is defined as the temperature of a blackbody with the same average
frequency as the observed data. For a protostar to be a Class 0 source, it
must have either T$_{bol}$ $\leq$ 70 K or L$_{submm}$/L$_{bol}$ $>$ 0.005 \citep{chen1995, andre1993}; however,
protostars can fulfill the L$_{submm}$/L$_{bol}$ criterion, but not necessarily the T$_{bol}$ criterion and 
we refer to these sources as Class 0/I sources.

L1157-mm (IRAS 20386+6751) is a Class 0 source with $T_{bol}$  = 40 K and L$_{submm}$/L$_{bol}$ = 0.018
and it has the lowest luminosity of the sample with $L_{bol}$ $\sim$ 4.3 $L_{\sun}$.
The envelope surrounding L1157-mm is also found to have a flattened/filamentary structure 
that is extended perpendicular
to the outflow direction \citep{looney2007}. Observations of the outflow indicate that 
the protostar is viewed with a near edge-on orientation \citep{gueth1996}. L1157-mm was
studied extensively with CARMA, aiming to detect emission from its disk \citep{chiang2012}; however, 
the source remained unresolved at a resolution of 0\farcs3 (90 AU) at 1.3 mm and 3.4 mm. 
Models of the dust emission without an unresolved disk component are a modestly 
better fit to the data than those with a disk, suggesting that any disk around L1157-mm
has very little overall mass; model-dependent estimates range from 0.004 $M_{\sun}$ to 
0.024 $M_{\sun}$. 

The protostar L1165-SMM1 (IRAS 22051+5848, HH 354 IRS) was first examined in the submillimeter by
\citet{visser2002} with 850 \micron\ dust continuum maps and outflow maps; 
\citet{reipurth1997a} had previously noted the Herbig-Haro object near this source. 
There is also a bright infrared source (IRAS 22051+5849) 89\arcsec\ ($\sim$0.13 pc) away 
from L1165-SMM1, but it is not detected in the submillimeter maps by \citet{visser2002}.
The \textit{Spitzer} c2d survey identified this source as a candidate young stellar object and it has
a spectral index consistent with a Class II source \citep{evans2009}. We classify
L1165-SMM1 as a Class 0/I source given that $L_{submm}$/$L_{bol}$ = 0.012 and 
T$_{bol}$ $\sim$ 78 K; this classification makes use of new \textit{Herschel} 
photometry that sample the protostellar SED between 100 \micron\ and 500 \micron\ (see Appendix).

The difference between the new and previous $T_{bol}$ determinations for L1165-SMM1 is due to the inclusion
of near-infrared fluxes in the calculation of $T_{bol}$ 
and a well-sampled SED out to the millimeter from \textit{Herschel}.
We note that inclination effects can lead to elevated values of $T_{bol}$, causing 
Class 0 sources to be classified as Class I sources. This is due to 
near to mid-infrared scattered light emission escaping through 
the outflow cavities \citep{jorgensen2009, launhardt2013} and the frequency-weighted nature
of $T_{bol}$ makes it very sensitive to the level of short-wavelength flux. $L_{submm}$/$L_{bol}$
is also affected by inclination, but less than $T_{bol}$ since it is the integral of the flux
density rather than a frequency-weighted average wavelength.

The protostar CB230 IRS1 (L1177, IRAS 21169+6804) resides within an isolated globule identified by \citet{cb1988}. 
Near-infrared imaging previously detected a companion separated from 
IRS1 by $\sim$10\arcsec\ (3000 AU) to the east \citep{yun1996}; we refer to this companion as IRS2.
IRS2 has been detected at wavelengths between 1.2 \micron\ and 24 \micron\ \citep{massi2008}, 
but is undetected in \textit{Herschel} data at 100 \micron\ \citep{launhardt2013}, 
SCUBA at 450 \micron\ \citep{launhardt2010}, and OVRO 1.3 mm and 
3.4 mm data \citep{launhardt2001a,launhardt2001b}.
CB230 IRS1 is also classified as a Class 0/I source with 
$L_{submm}$/$L_{bol}$ = 0.037 and $T_{bol}$ = 189 K \citep{launhardt2013}.

Both L1165-SMM1 and CB230 IRS1 are found to be FU Ori-like
objects from CO and water absorption in their near-infrared spectra 
\citep{reipurth1997b,green2008,massi2008}. The distances to all the
sources are uncertain given their association with the Cepheus Flare, 
but are estimated to be between 250 pc and 440 pc \citep{viotti1969,straizys1992,kun1998,kun2008};
we have adopted a distance of 300 pc as we have in previous studies \citep{tobin2010a,tobin2011}.
The selected protostars all have rather filamentary envelopes 
in dust extinction at 8 \micron\ \citep{tobin2010a} and have strong emission in the dense gas tracers \nht\ and \nthp.
All the protostars have velocity gradients normal to the outflow direction that may be indicative of rotation
or infalling flows along the envelope; L1165 and CB230 IRS1 have gradients that 
are 2$\times$ larger than the gradient in  L1157-mm \citep{tobin2011}.

\subsection{VLA Observations}
The observations were taken in the standard continuum mode with 2 GHz of total bandwidth covered by 16, 128 MHz wide sub-bands
that are further divided into 64, 2 MHz channels with four polarization products. Details of the observations are
listed in Table 2. The observations in C configuration
were taken between 2012 February and March. Each track was shared between the
three target sources, with alternating observations on a target source for 4 minutes and 
then observing the gain calibrator J2022+6136 for 1 minute. An additional C-array observation was
taken in Jul 2013 at 4.0/6.5 cm for L1165 and CB230 to complement the shorter wavelength data. The observations
at 4.0/6.5 cm observed the calibrator J2148+6107 for 1 minute and then each source for $\sim$12 minutes. In the 4.0/6.5 cm
observations, the two 1 GHz basebands were separated by 1.8 GHz with central wavelengths of 6.5 cm and 4.0 cm in order
to measure a spectral index from the 4.0/6.5 data alone.
The observations in B configuration were carried out in 3 hour
scheduling blocks for each source in 2012 June. The protostars were observed at $\lambda$ = 7.3 mm, 1.4 cm, and 3.3 cm during the
same track. The observations were taken in a fast switching mode to compensate for rapid atmospheric
phase fluctuations, observing the source for 1 minute and the gain calibrator for 30 seconds. The lack of a strong 
calibrator less than 10\degr\ away from any of the protostars reduced the observing efficiency to $\sim$25\% 
due to slew times. The A configuration observations were taken in 1 hour blocks over many dates between October 2012 and 
December 2012 (Table 2). The observations at 7.3 mm, 1.4 cm, and 3.3 cm were taken in separate
observing blocks to maximize probability of scheduling; 1.4 cm and 3.3 cm data can be observed during more marginal weather
than 7.3 mm. The same gain calibrators were used
for both A and B configurations and either 3C48 or 3C286 was used for flux density calibration. For all configurations,
we did an interferometric pointing scan at 3.3 cm on the primary calibrator every hour 
for the 7.3 mm and 1.4 cm observations, as pointing errors become apparent on this timescale; 
pointing is not necessary at 3.3 cm or 4.0/6.5 cm.

Each dataset was reduced using the Common Astronomy Software Application (CASA)\footnote{http://casa.nrao.edu}.
The raw visibility data were downloaded from the VLA archive as CASA measurement sets.
We first inspected the amplitudes
and phases in each dataset and flagged uncalibrateable data, characterized by low amplitudes, amplitude jumps, 
phase jumps, and/or periods of high phase decorrelation. The amplitudes across all spectral windows were also 
inspected for issues such as radio frequency interference. Following the data editing, we used the \textit{setjy}
task with the appropriate clean component model to set the absolute flux density scale of the flux density calibrators identified in Table 2.
We then checked for updated antenna positions and generated a gain table to correct the phases using the \textit{gencal} task.
The atmospheric opacity was determined from the weather station data via the \textit{plotWeather} task and the opacity in each 
spectral window was used as input in all subsequent amplitude calibration tasks.
We then performed first-pass gain calibration for amplitude and phase on the flux calibrator and primary
gain calibrators using the \textit{gaincal} task. When 3C48 was used as the flux calibrator, we only 
considered baselines with uv-distances less than 1000 k$\lambda$ since the source is resolved-out on longer baselines.
We then used the \textit{fluxscale} task to determine the flux of the primary
gain calibrators at each observed wavelength. The \textit{setjy} task was then used to set the measured fluxes of the gain
calibrators. The absolute calibration uncertainty is expected to be 10\% and we only quote statistical uncertainties hereafter.

We next calibrated the bandpass solution using the \textit{bandpass} task and we used the primary gain calibrator
to determine the bandpass solution. Therefore, it was necessary to calculate a gain 
solution on 10 second timescales to compensate for the rapid phase variations 
over the course of the track, prior to calculating the bandpass solution. 
After bandpass calibration, we performed a final gain calibration
for amplitude and phase using the \textit{gaincal} task, over the entire length of each calibrator scan.
The calibration tables from the bandpass and gain calibrations were then applied to the data
using the \textit{applycal} task. Following the application of the calibration data, we verified that the calibrations
were properly applied by examining the phase and amplitudes of the calibrated data. 
The B and C configuration data were reduced using
CASA 3.3, the A configuration data were reduced using CASA 3.4 and 4.0, and the 4.0/6.5 C configuration data were
reduced using CASA 4.1. No significant
differences were noticed between reductions with different CASA versions.

Following the gain calibration, the data at each wavelength were imaged using the \textit{clean} task. The measurement sets
for each configuration and track were combined using the \textit{concat} task and we deconvolved each dataset
as a whole. We used the Hogbom
clean algorithm because this method tends to work better than the standard Clark algorithm for 
snapshot observations with a sparsely sampled uv-plane\footnote{CASA Cookbook}. We used a clean threshold 
of 1.5$\times$ the noise measured from the dirty map and a clean mask drawn closely around the areas of source emission. 
We also imaged the sources with uv-tapering at 500 k$\lambda$, 1000 k$\lambda$, 1500 k$\lambda$, and 2000 k$\lambda$ to bring out
more large-scale structures.

The resulting images from the VLA at 7.3 mm in A configuration have an 
angular resolution up to $\sim$0\farcs06 ($\sim$18AU; natural-weighting) and
are the highest resolution images obtained toward these particular protostars. The 1.4 cm and 3.3 cm data have
corresponding angular resolutions of $\sim$0\farcs2 ($\sim$60AU) and $\sim$0\farcs3 ($\sim$90AU); the exact resolutions
depend on the value of the robust parameter and any applied tapering. The 4.0 cm and 6.5 cm data had resolutions
of $\sim$4\arcsec\ and 7\arcsec\ respectively. In the case of the
7.3 mm and 1.4 cm data, we generally applied tapering and use a robust parameter of 2 which is equivalent to natural weighting.
For the 3.3 cm data, we used robust values of 0.5 and 1 for L1165-SMM1 and CB230 IRS1 
respectively to achieve higher resolution. 

\subsection{CARMA Observations}

L1165-SMM1 was observed by CARMA in C and B configurations at 1.3 mm in December 2012 and January 2013, details are given in Table 3. 
The B-array data were taken during exceptionally good weather as indicated by the low opacity at 225 GHz. The 
local oscillator frequency was 225.05 GHz and 6 correlator windows were configured to have 500 MHz bandwidth
 for continuum observation, yielding a total continuum bandwidth of 6 GHz (dual-side band)
 and the remaining two windows were configured to have 31 MHz bandwidth and 0.13 \kms\ channels.
 These two windows were set to observe $^{12}$CO, $^{13}$CO, and C$^{18}$O 
($J=2\rightarrow1$), with $^{12}$CO and C$^{18}$O being observed in the same 
spectral window in opposite side-bands. The data were taken in a standard observing
loop, science target observations bracketed by gain calibrator observations. 
The C configuration data observed the calibrator
3C418 for 3 minutes and then the source for 9 minutes; the first B 
configuration track observed 3C418 for 2 minutes
and then L1165-SMM1 for 9 minutes. The second B configuration track 
observed 3C418 for 2 minutes, L1165-SMM1 for 8 minutes
and then a test point source 2148+611 for 1 minute.

CB230 IRS1 was observed by CARMA in C configuration at 3.4 mm in January 2013 and May 2013. 
The local oscillator frequency was 90.9 GHz and one 500 MHz window was configured for continuum
observation yielding 1 GHz of continuum bandwidth (dual-side band). Spectral lines were observed
in the remaining windows, but these data will not be considered here.
The data were taken in a standard observing loop, science target observations bracketed 
by gain calibrator observations. The calibrator (1927+734) was observed for 3 minutes and 
then the CB230 IRS1 for 15 minutes. 

The data were reduced with the MIRIAD software package \citep{sault1995}. We first split out the noise source observations
and then applied corrected baseline solutions when needed using the \textit{uvedit} task. 
Following this, we applied calibrations for transmission line
length changes due to thermal expansion using the \textit{linecal} task. After these initial calibrations, we inspected the 
phase and amplitudes of the data versus time and flagged obvious bad data, consisting of but not limited to phase jumps, high
phase variance, and low amplitudes. We then applied bandpass corrections determined with the \textit{mfcal} task,
using the noise source to correct the 31 MHz bands and a bright quasar for the 500 MHz bands. Following 
these calibrations, we determined the flux density of the primary gain calibrator
using the absolute flux calibrators listed in Table 3 with the \textit{bootflux} 
task. The flux calibration uncertainty is estimated to be 10 - 20\% and we only quote statistical uncertainties hereafter.
 We then calibrated the complex gain and phases on the wide-band continuum 
data using the \textit{mfcal} task and transferred these solutions to the 31 MHz bands. We then ran \textit{mfcal}
on the 31 MHz windows to correct for phase offsets between the 500 MHz and 31 MHz bands.

The data were imaged by first inverting the visibility data to construct the dirty map with the \textit{invert} task. The 
dirty map was CLEANed with the \textit{mossdi} task and the clean components are then convolved with the clean beam and added
back to the residuals using the \textit{restor} task to output the final clean map.

The image fidelity in B-array was verified by imaging the faint secondary calibrator taken in the second B-array track, finding
an unresolved point-source. While we did not observe a secondary calibrator in the first track, the science target structure
was consistent between the first and second track, indicating that the conditions were similarly good.

\subsection{Ancillary Infrared Data}

Near and mid-infrared data are also included in this paper for CB230 to illustrate the 
larger-scale structure of CB230 IRS1. The Ks-band near-infrared
imaging of CB230 IRS1 was taken with the 2.4 m Hiltner telescope of the MDM Observatory at Kitt Peak on 2008 June 12. The data
were taken in a 5-point dither pattern and off-source sky frames were observed every 6 minutes. The data were reduced
using standard methods for near-infrared data reduction in IRAF\footnote{IRAF is distributed by the National Optical Astronomy Observatories,
which are operated by the Association of Universities for Research
in Astronomy, Inc., under cooperative agreement with the National
Science Foundation.}. See \citet{tobin2008} for more details of the observations and reduction procedure.

The \textit{Spitzer} IRAC and MIPS photometry were published in \citet{massi2008}, but they did not
show any \textit{Spitzer} images in their work. We previously published the 3.6 \micron\ image 
in \citet{tobin2010a} and the reduction procedure can be found there; the MIPS 24 \micron\ image is taken
directly from the \textit{Spitzer} archive.

\section{Results}

The VLA imaging toward the protostars L1165-SMM1, L1157-mm, and CB230 IRS1
has enabled us to examine their circumstellar structure in detail 
previously unattainable with the old VLA system and at resolutions currently unmatched 
by a millimeter interferometer. The emission at 7.3 mm is expected to be dominated by
thermal dust emission with a contribution of optically thin free-free emission from the protostellar
jets \citep{anglada1995,anglada1998, shirley2007}; the 1.4 cm emission is expected to have roughly
equal contributions from dust emission and free-free, and the 3.3 cm, 4.0 cm, and 6.5 cm data are expected to be dominated by free-free
emission.

\subsection{L1165-SMM1}

\begin{figure}[!ht]
\begin{center}
\includegraphics[angle=-90, scale=0.75]{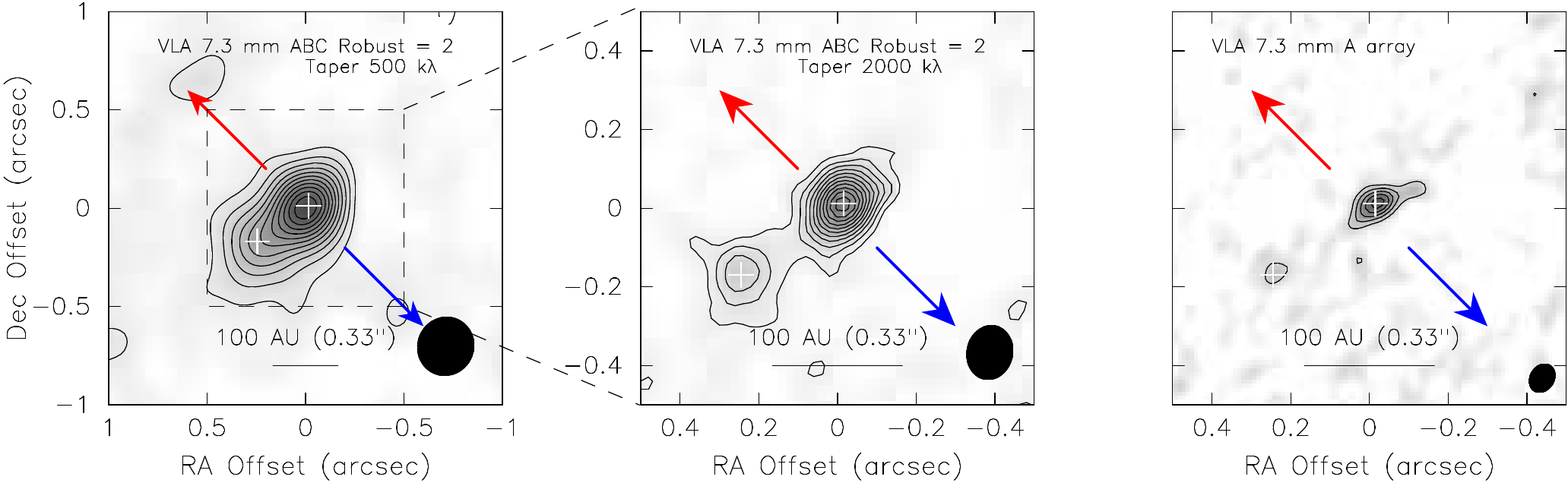}

\end{center}
\caption{L1165-SMM1 images at 7.3 mm at resolutions of $\sim$0\farcs3 (left), 0\farcs1 (middle), and 0\farcs07 (right).
Notice the extended nature of L1165-SMM1 at low resolution, splitting into
multiple sources at high resolution. The primary appears extended perpendicular to the outflow 
in the 2000 k$\lambda$ tapered image and the A-array image, consistent with a compact circumstellar disk.
Contours start at $\pm$3$\sigma$ and increase in 2$\sigma$ intervals, where
$\sigma$= 28.5 $\mu$Jy beam$^{-1}$, 21.3 $\mu$Jy beam$^{-1}$, and 28 $\mu$Jy beam$^{-1}$for the 500 k$\lambda$ tapered image,
2000 k$\lambda$ tapered image, and A-array-only image respectively. The size of the beam in each image is drawn in the lower right. }
\label{L1165-7mm}
\end{figure}

The 7.3 mm data for L1165-SMM1 are shown in Figure \ref{L1165-7mm}; the left panel shows 
the larger-scale ($\sim$150 AU) emission with the visibilities tapered at 
500 k$\lambda$ yielding $\sim$0\farcs3 resolution. The emission on 0\farcs3 scales
is clearly asymmetric with extension toward the southeast. The 2000 k$\lambda$ tapered
image, having a resolution of $\sim$0\farcs1, is shown in the middle panel of Figure \ref{L1165-7mm},
and the source is resolved into an apparent binary
system at the higher resolution, with a separation of 0\farcs3 ($\sim$100 AU). 
Both sources are clear detections, but the secondary is about
3$\times$ fainter than the primary at 7.3 mm. The primary 
also appears marginally resolved in the central panel of Figure \ref{L1165-7mm} and the companion  
is unresolved. Finally, the highest resolution image with A-array data only, is shown in the
right panel of Figure \ref{L1165-7mm}. This image shows that the primary source is resolved, appearing disk-like ($\sim$60 AU diameter)
and extended perpendicular to the outflow; both sources appear fainter due to more emission being resolved-out in the A-array only data.

\begin{figure}[!ht]
\begin{center}
\includegraphics[angle=-90, scale=0.3]{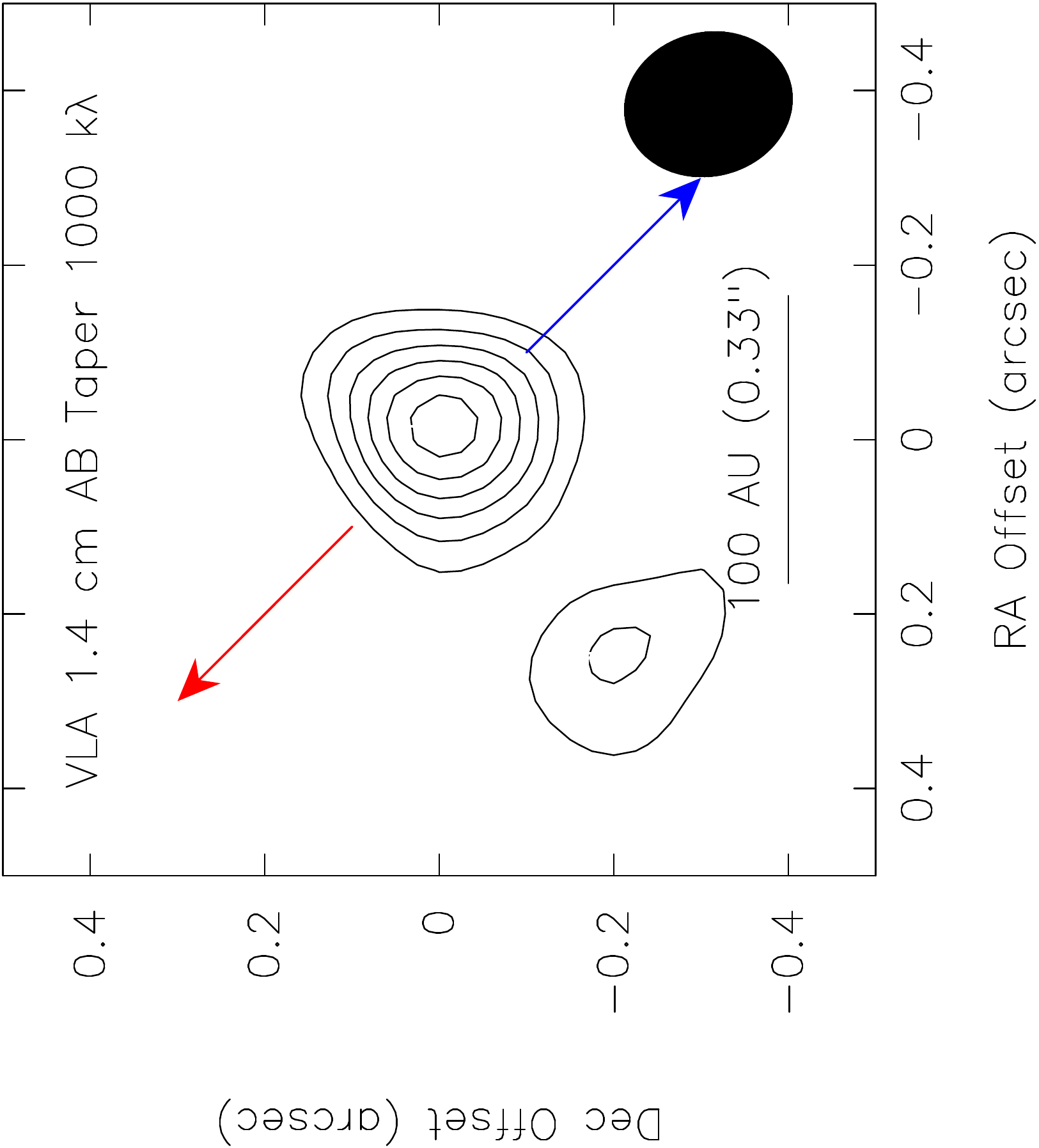}
\includegraphics[angle=-90, scale=0.3]{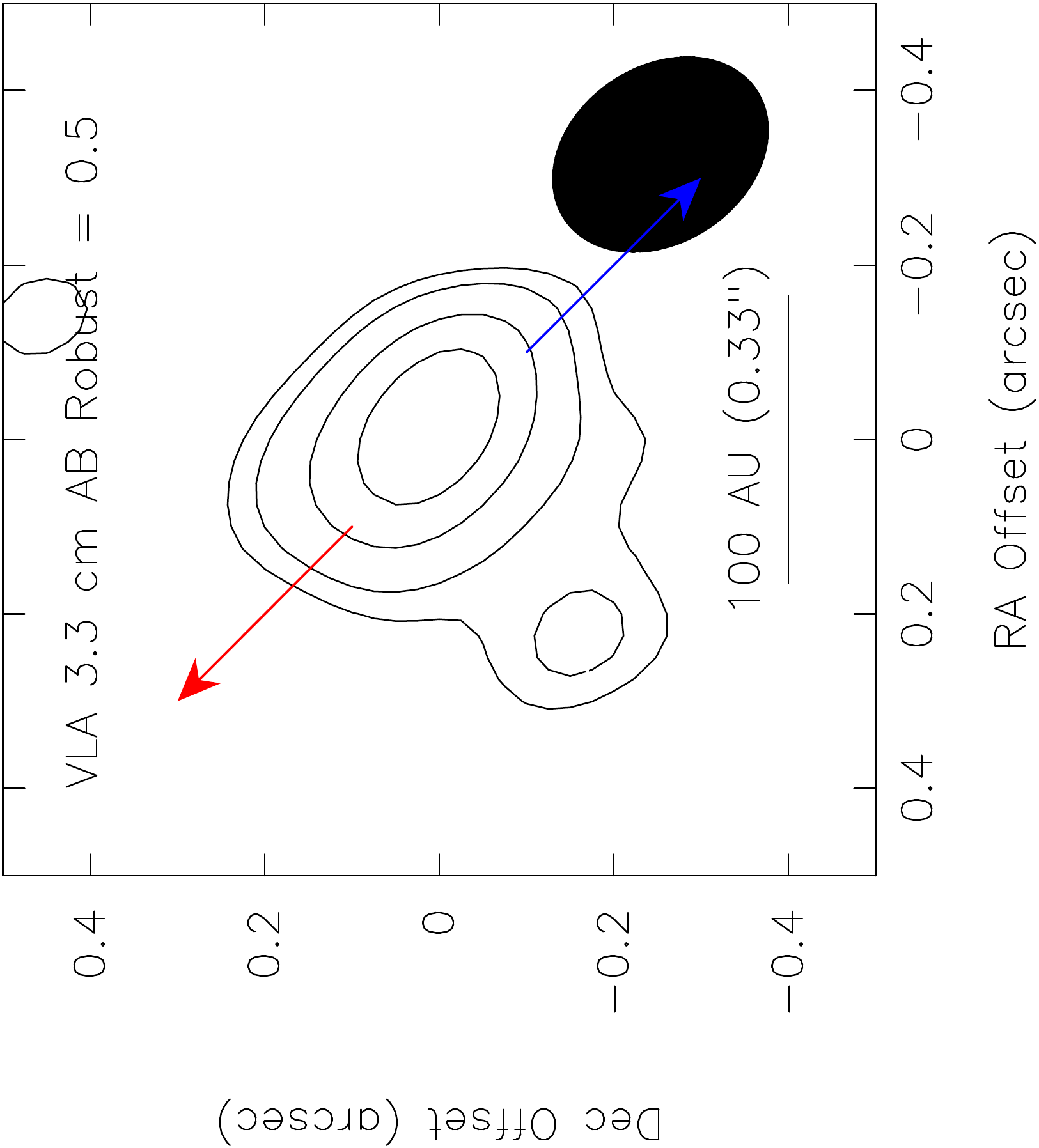}
\end{center}
\caption{L1165-SMM1 images at 1.4 cm and 3.3 cm. Both sources are clearly detected at centimeter wavelengths,
indicative of free-free emission from the thermal jets of each source. The centroids of the secondary
are slightly offset at each wavelength, which may result from the free-free emission not being 
centered at the dust continuum peak at 7.3 mm and blending with the primary in the 3.3 cm image.
Contours start at $\pm$3$\sigma$ and increase at 2$\sigma$ intervals in the 1.4 cm image where $\sigma$ = 12.8 $\mu$Jy beam$^{-1}$ and 
in the 3.3 cm image the contours are [2, 3, 5, 7, 9, ...] $\times$ $\sigma$, where $\sigma$ = 9.6 $\mu$Jy beam$^{-1}$.}
\label{L1165-14mm}
\end{figure}

The centimeter-wave emission (1.4 cm and 3.3 cm) is shown in Figure \ref{L1165-14mm}.
 The 1.4 cm image shows that both sources are point-like and distinct 
from each other at this wavelength. At 3.3 cm, the primary source is well-detected, but the 
secondary is only a 3.5$\sigma$ detection and blended with the primary. 
The detection of the secondary source at both 1.4 cm and 3.3 cm
wavelengths is evidence that it is a bonafide protostar, given the large contribution from free-free emission at
these wavelengths requires some source of ionization. Moreover, since
the secondary is located normal to the outflow direction of the primary, it is unlikely that an
outflow shock could be causing a false secondary in dust continuum and/or free-free emission. We examined the position
of the young stellar object (IRAS 22051+5849) located 89\arcsec\ north at 3.3 cm, 4.0 cm, and 6.5 cm image 
and did not find evidence of emission. We do not show the 4.0 cm or 6.5 cm data for the sake of brevity, but list the flux densities in Table 4.

\begin{figure}[!ht]
\begin{center}
\includegraphics[angle=-90, scale=0.4]{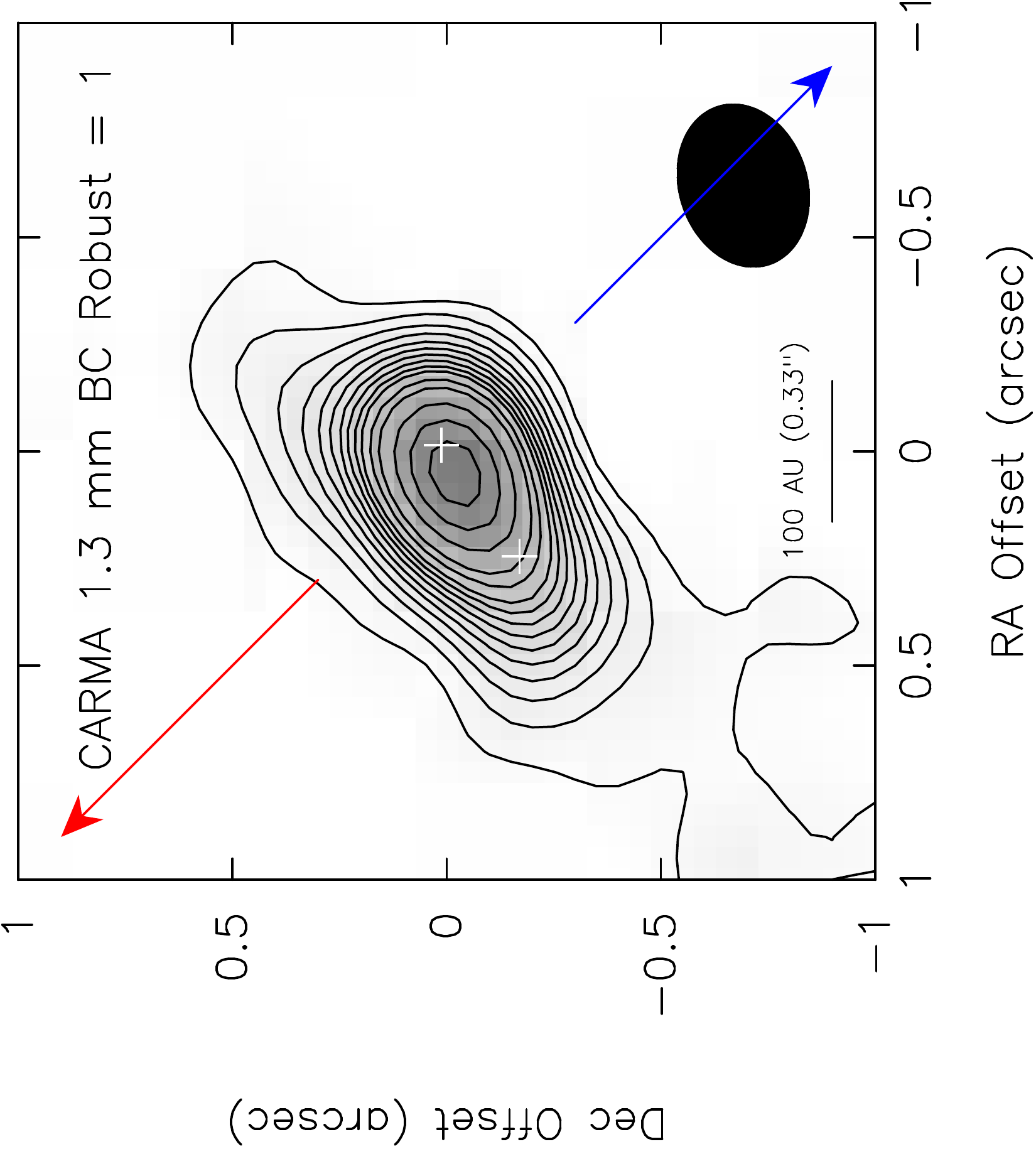}
\includegraphics[angle=-90, scale=0.4]{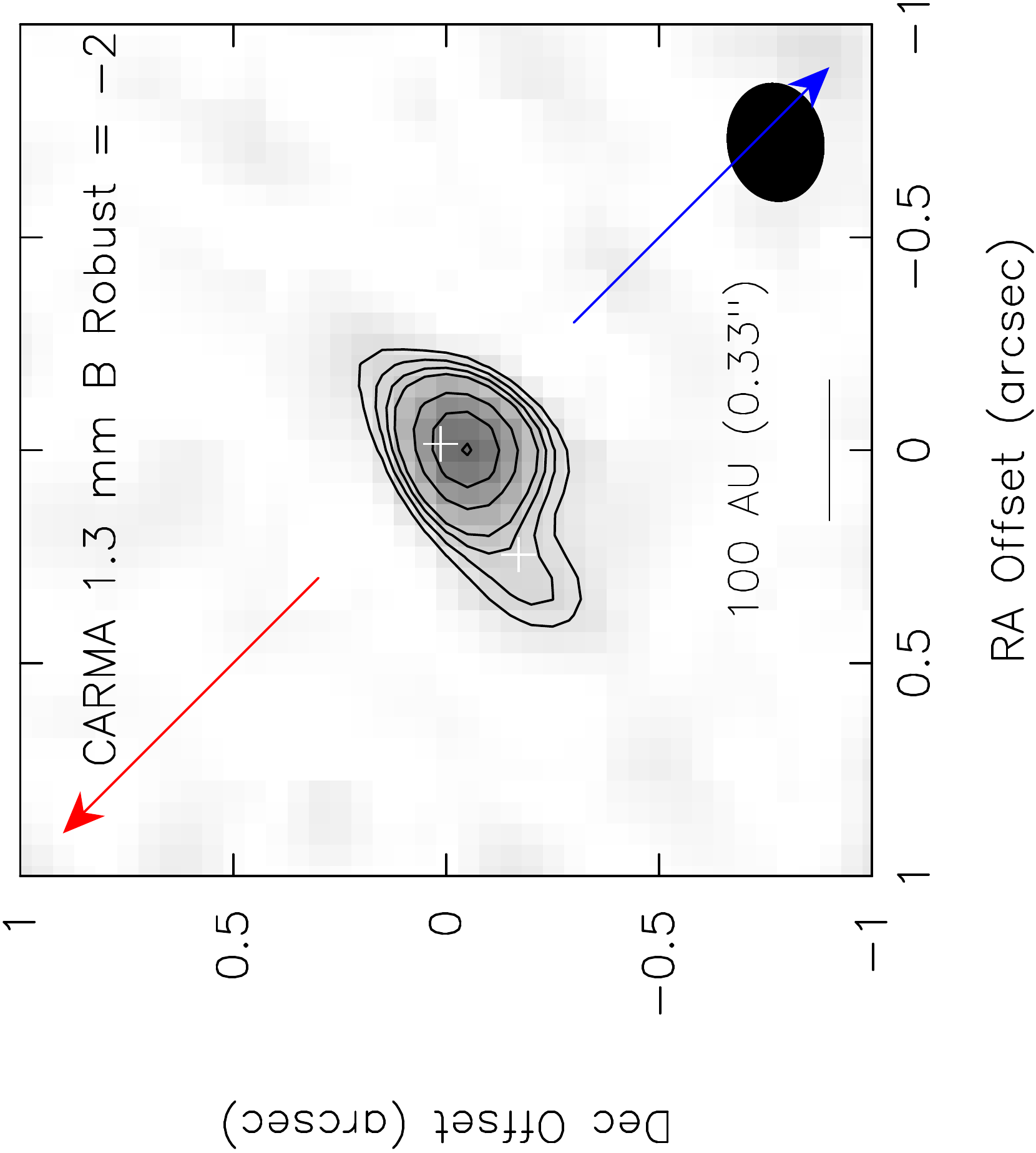}
\includegraphics[angle=-90, scale=0.4]{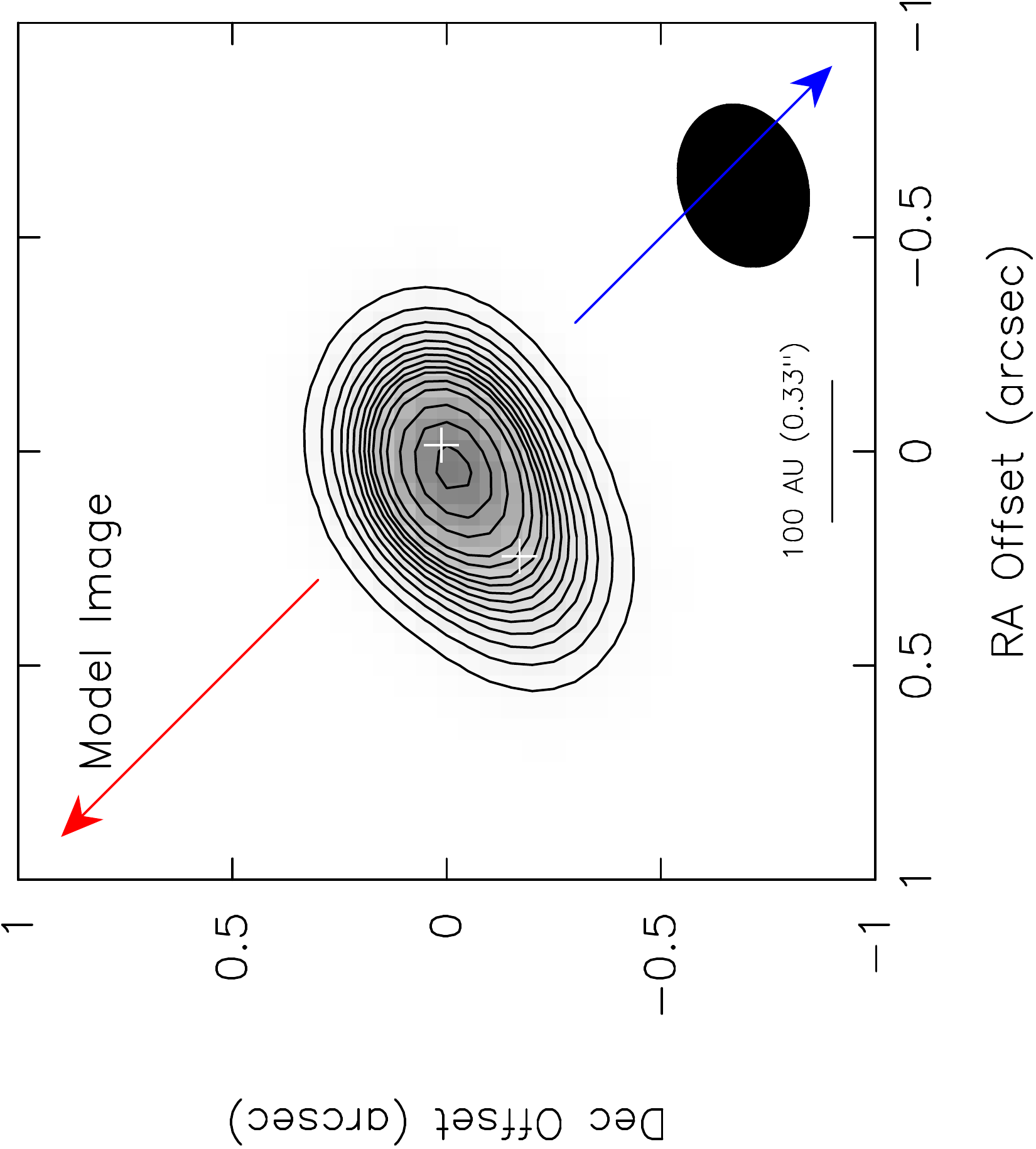}
\includegraphics[angle=-90, scale=0.4]{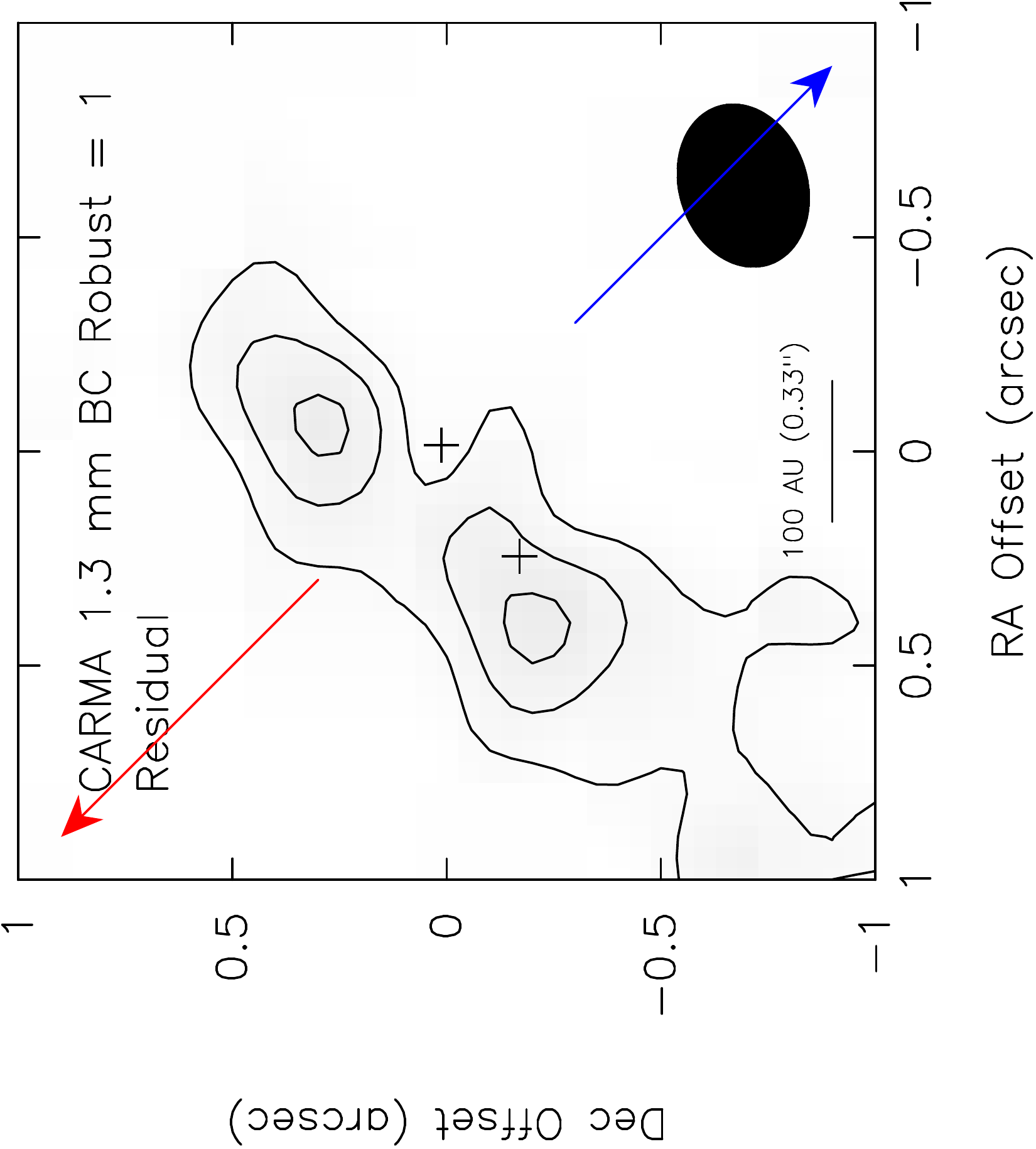}
\end{center}
\caption{L1165-SMM1 images at 1.3 mm showing an extended structure 
normal to the outflow direction and surrounding the two
sources. The top left panel is the combined B and C configuration image, showing more extended
structure than the 7.3 mm image with 500 k$\lambda$ tapering (Figure \ref{L1165-7mm}), with a robust
weighting of 1. The top right 
panel shows the B configuration image with a robust weighting of -2, yielding higher 
resolution and less extended emission. This image finds a strong continuum source 
peaked near the 7.3 mm peak of the primary protostar with a weaker extension toward the
position of the secondary. The bottom left panel is a model image of the two blended
point sources with the same resolution as the top left image. The bottom right panel
is the residual of the top left image after subtraction of the model image. The residual image
shows an extended structure perpendicular to the outflow, consistent with a possible circumbinary disk.
The contours plotted in the top left, bottom left and bottom right panels (robust =1 images)
 are [-3, 3, 6, 9, 12, 15, 18, 21, 24, 27, 30, 35, 40, 50, 60, 70] $\times$ $\sigma$ 
where $\sigma$= 0.69 mJy beam$^{-1}$.
The top right panel (robust =-2 image) contours are [-3, 3, 4, 5, 6, 9, 12, 15] $\times$ $\sigma$ 
where $\sigma$= 2.9 mJy beam$^{-1}$. }
\label{L1165-1mm}
\end{figure}

Follow-up observations of dust continuum emission at 1.3 mm were obtained with 
CARMA, having $\sim$0\farcs3 resolution. The 1.3 mm data are shown in 
Figure \ref{L1165-1mm} and trace an extended, flattened structure
oriented perpendicular to the outflow, possibly a circumbinary disk.
The disk structure appears to be surrounding the two protostars and it appears
to be more extended southeast in the 1.3 mm data than at 7.3 mm. There is also
an extension toward the northwest in the 1.3 mm data not seen at 
7.3 mm. We also found that when the 1.3 mm data
were imaged with the robust parameter set to -2, increasing the 
resolution of our deconvolved image (with increased
noise), the two sources are still blended and there
is extended structure associated with the primary (Figure \ref{L1165-1mm}, top right 
panel). Gaussian fitting enabled the sources to be separated; the flux density of the 
secondary is a factor of $\sim$5 less than the primary, a larger ratio than at 7.3 mm. 
The integrated flux densities measured with CARMA 
and the VLA are given in Table 4 as well as Gaussian fits to the image data.

In order to better determine the properties of the circumbinary structure
around L1165-SMM1, it is necessary to decompose its emission from that 
which is more closely associated with the two protostars. 
We used the positions of the two sources from the VLA 7.3 mm map as a prior to 
fit the flux densities of the sources in the 1.3 mm robust = -2 map.
Point sources rather than 
Gaussians were used in this case because our goal is to 
separate the extended structure from the compact structure.
We performed this step on the robust = -2 map because it has less extended structure and better
isolates the emission from the individual sources.
Next, we used the positions and fitted flux 
densities (48.1 mJy and 13.7 mJy) to construct a model image
of the two blended point sources at the resolution of the robust = 1 map 
(Figure \ref{L1165-1mm}, bottom left). We then subtracted the model 
from the data and the residual image is shown in the bottom right panel of 
Figure \ref{L1165-1mm}; the flux density in the residual image is 20.5$\pm$3 mJy. 
The residual image shows that the robust = 1 map cannot be reproduced by two blended point
sources, having what appears to be a more extended circumbinary structure. 
We note that this structure is slightly offset to the northeast from the primary source; however, 
this could be due to projection effects because L1165-SMM1 is not viewed edge-on. While
this analysis was done in the image plane, we also performed the same analysis in the uv-plane
by subtracting point sources with the \textit{uvmodel} task and obtained a similar result.

\begin{figure}[!ht]
\begin{center}
\includegraphics[scale=0.7]{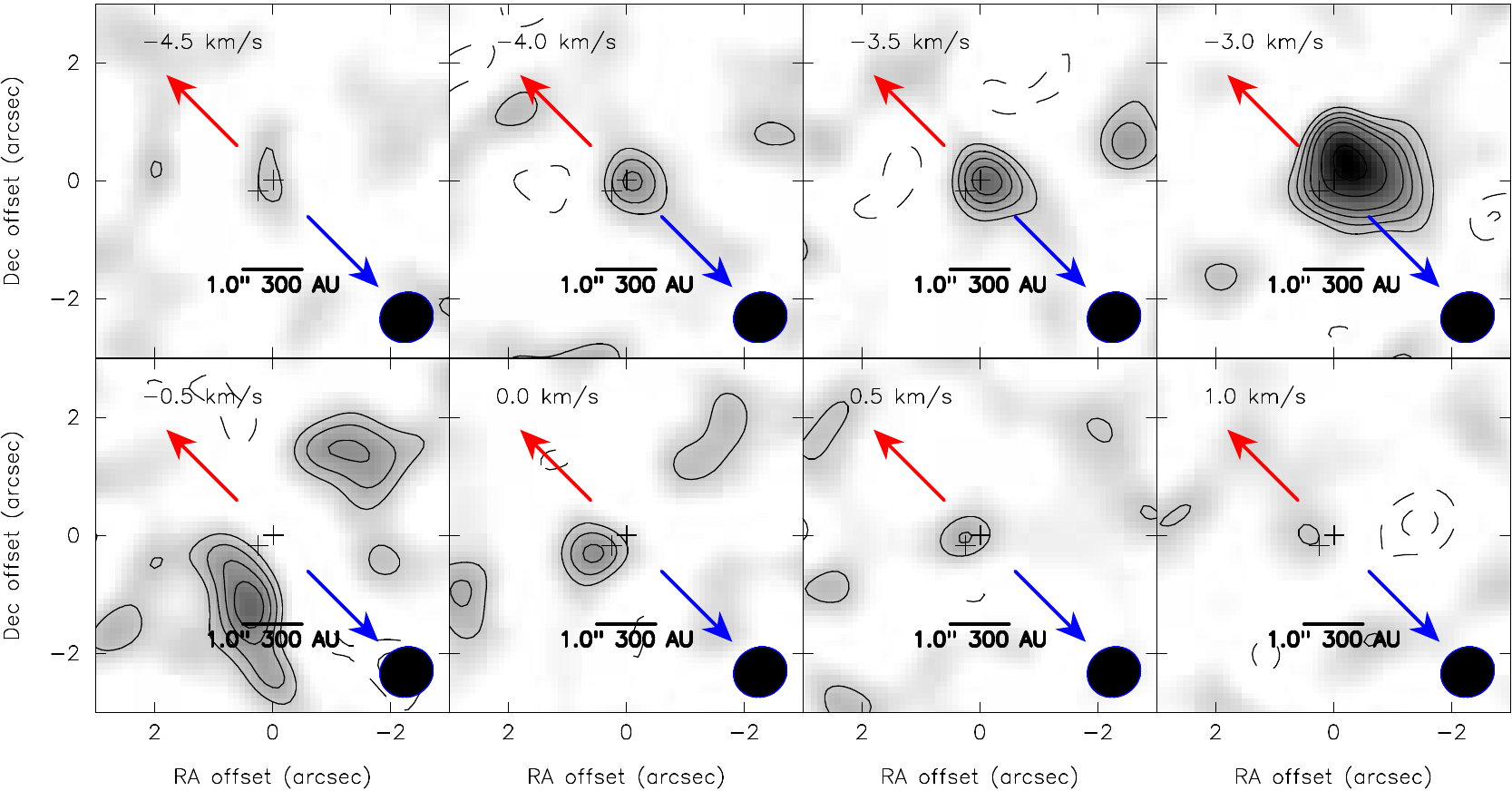}
\end{center}

\caption{$^{13}$CO channel maps for L1165-SMM1, showing evidence 
for rotation on the scale of the disk; contours start at $\pm$2$\sigma$ and increase
by 1$\sigma$; the systemic velocity is $\sim$-1.5 \kms. The noise level in 
the images is 80 mJy/beam with 0.5 \kms\ channels.}
\label{L1165-13CO}
\end{figure}

\begin{figure}[!ht]
\begin{center}
\includegraphics[angle=-90, scale=0.7]{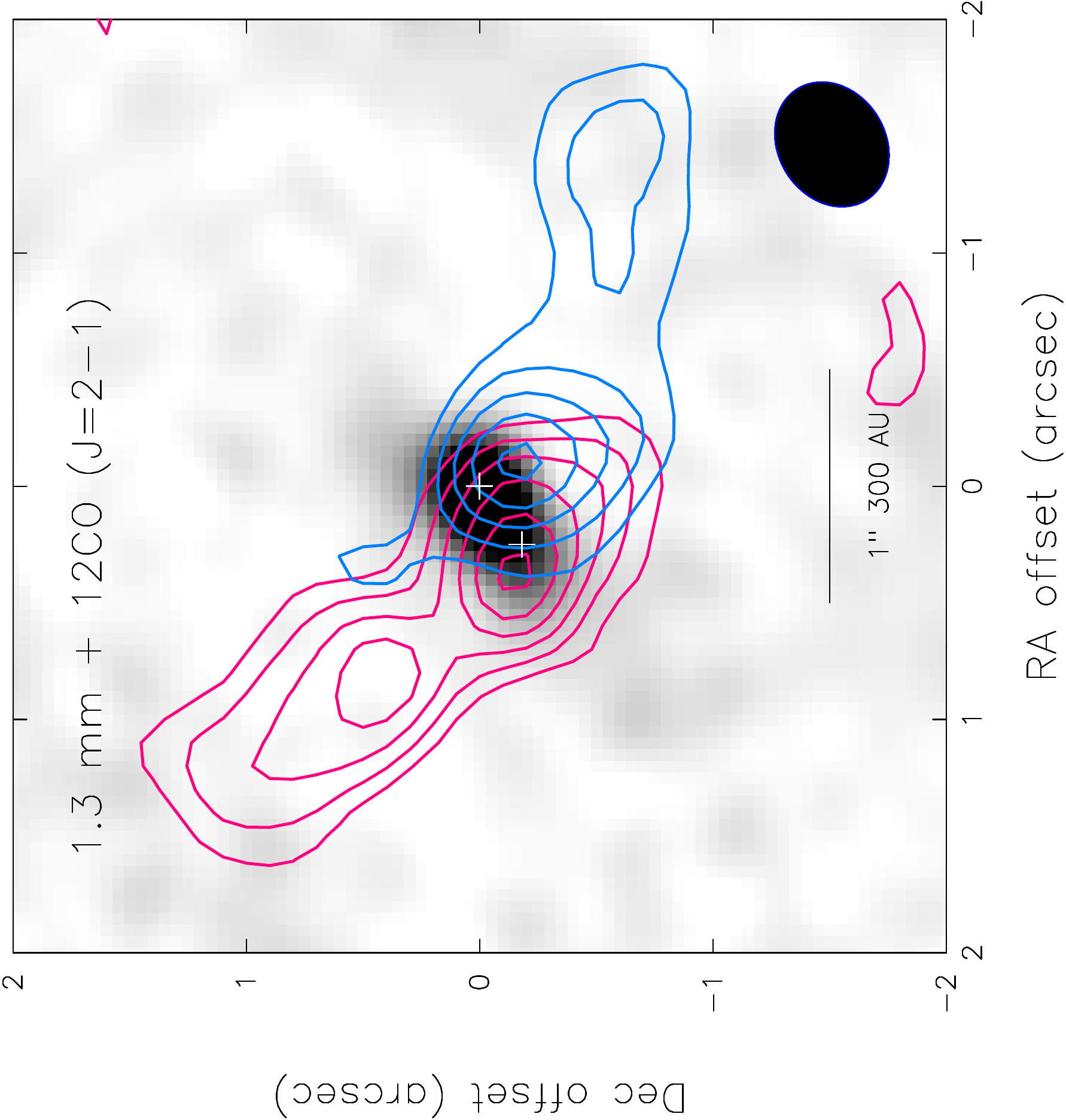}

\end{center}
\caption{L1165-SMM1 dust continuum at 1.3 mm (grayscale) with $^{12}$CO blue and redshifted emission overlaid tracing the molecular
outflow from L1165-SMM1. Notice that the blueshifted emission seems to originate from the
primary source, while the red shifted emission appears
to originate from the companion. The redshifted emission is integrated between -0.5 \kms\ and 
2.5 \kms\ and the blue shifted emission is integrated between -8.0 \kms\ and -4.5 \kms.
The contours start at $\pm$5$\sigma$ and increase in increments of 2$\sigma$; $\sigma$=5.67 K \kms\ for 
the red and $\sigma$=6.07 K \kms\ for the blue.}
\label{L1165-CO}
\end{figure}

We do find evidence for rotation on the scale of the circumbinary disk structure
in $^{13}$CO ($J=2\rightarrow1$) emission. Figure \ref{L1165-13CO} shows the
$^{13}$CO channel maps and the emission is
consistent with rotation on the scale of the disk. However, the signal-to-noise
of the data are not good enough to attempt a mass measurement or verify a Keplerian
velocity profile. The $^{12}$CO ($J=2\rightarrow1$) line was also 
detected in our CARMA observations, the integrated red and blueshifted
CO emission is shown in Figure \ref{L1165-CO}. The blue and red-shifted 
sides of these outflows are consistent with the outflow 
observations by \citet{visser2002}. While not well-resolved, the blue-shifted emission
is centered toward the primary source and the redshifted emission is centered near the secondary
source. The centroid shift between the blue and red-shifted $^{12}$CO emission
could have a number of causes: rotation of the disk, rotation of the outflow, 
or two outflows originating from either source.
The offset in the blue and red-shifted $^{12}$CO emission appears similar 
to that of CB26. The characteristics of CB26 could be explained as outflow rotation via the outflow being
coupled to the rotating disk; however, a misaligned
outflow from a binary companion was also suggested as a possibility \citep{launhardt2009}.

\subsection{CB230 IRS1}

\begin{figure}[!ht]
\begin{center}
\includegraphics[angle=-90, scale=0.75]{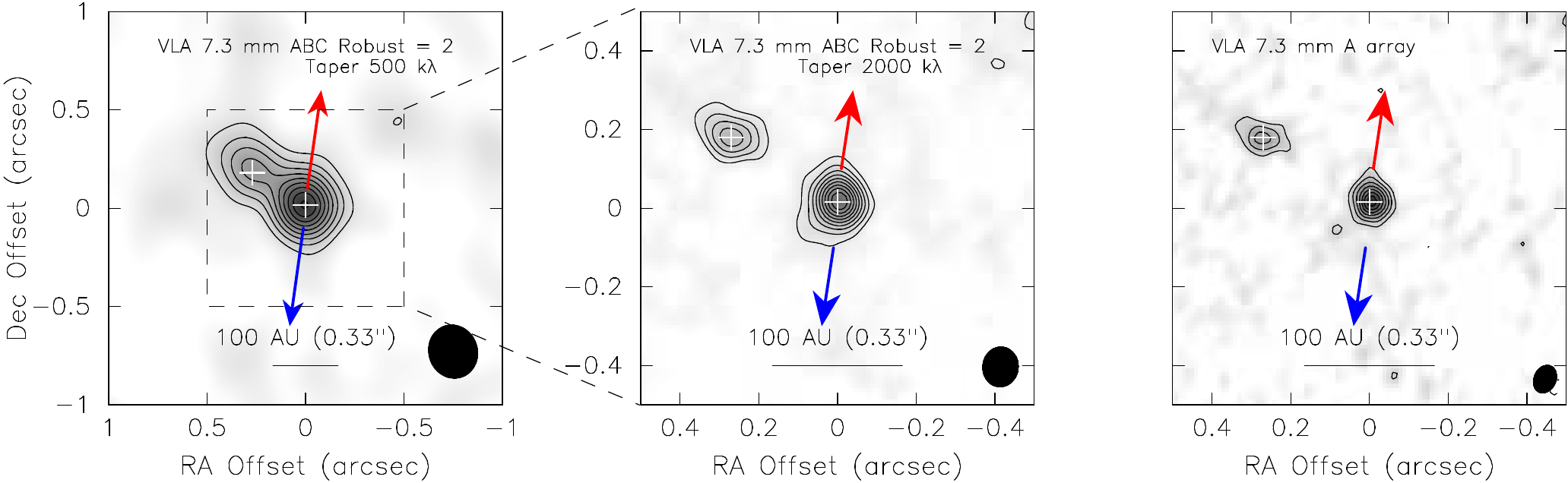}

\end{center}
\caption{CB230 IRS1 images at 7.3 mm. The sources appear marginally resolved in the lower
resolution image with 500 k$\lambda$ tapering and are quite distinct in the two higher
resolution images shown in the middle and right panels. The primary source appears
unresolved; however, the companion source appears extended roughly perpendicular
to the outflow direction. Contours start at $\pm$3$\sigma$ and increase in 2$\sigma$ intervals for
all images, where $\sigma$= 41.3 $\mu$Jy beam$^{-1}$, 27.5 $\mu$Jy beam$^{-1}$,
and 25.8 $\mu$Jy beam$^{-1}$ for the 500 k$\lambda$ tapered image,
2000 k$\lambda$ tapered image, and A-array-only image respectively.}
\label{CB230-7mm}
\end{figure}

The 7.3 mm data for CB230 IRS1 are shown in Figure \ref{CB230-7mm} with the combined A, B, and C 
configuration images shown with tapering at 500 k$\lambda$ and 2000 k$\lambda$ in
the left and middle panels; A-array only data are shown in the right panel.
A second source, separated from the main protostar 
by 0\farcs3 (100 AU) is detected in the two highest resolution images. 
The sources are marginally resolved at 0\farcs27 resolution (500k$\lambda$ tapering)
and resolved into two sources in the higher resolution data. The 
primary source appears unresolved in the high-resolution images; Gaussian fits
to the images, listed in Table 5, indicate that the source is only marginally
more extended than the beam. The secondary, however, appears to be 
resolved in both the A-array only and 2000 k$\lambda$ tapered images 
and it has a diameter of $\sim$60 AU and may be indicative of a compact circumstellar disk around the source.

\begin{figure}[!ht]
\begin{center}
\includegraphics[angle=-90, scale=0.25]{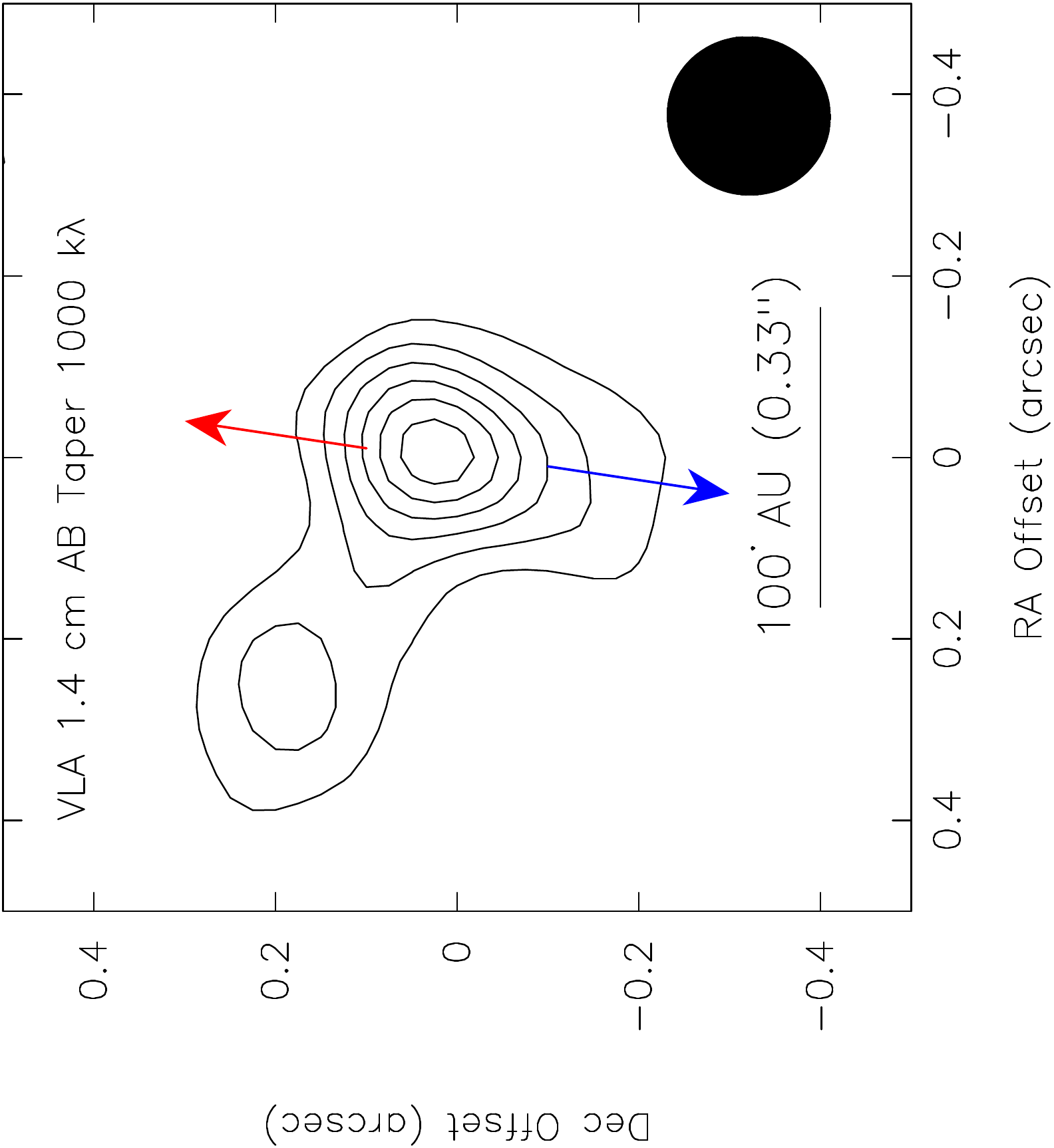}
\includegraphics[angle=-90, scale=0.25]{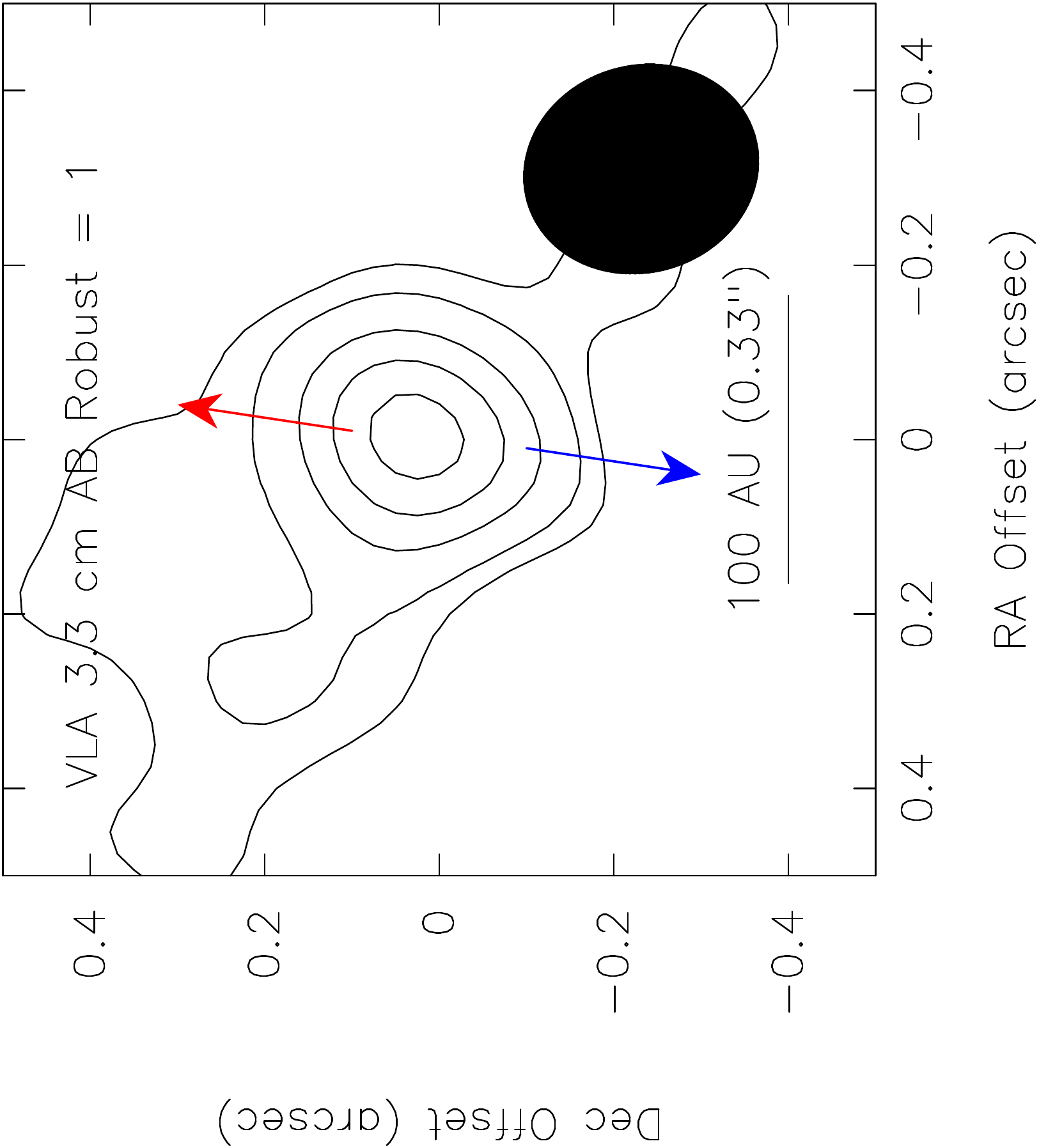}
\end{center}
\caption{CB230 IRS1 images at 1.4 cm (left) and 3.3 cm (right). The two sources are resolved at 1.4 cm, the
extended feature between them may simply result from the sources being blended. At 3.6 cm,
the secondary has a marginal peak above the 3 $\sigma$ level, but is surrounded by extended
emission at the 2$\sigma$ level. Contours start at $\pm$3$\sigma$ and increase 
at 2$\sigma$ intervals in the 1.4 cm image,  where $\sigma$= 11.5 $\mu$Jy beam$^{-1}$.
In the 3.3 cm image, the contours are $\pm$[2, 3, 5, 7, 9, ...] $\times$ $\sigma$, 
where $\sigma$ = 5.3 $\mu$Jy beam$^{-1}$. We start at $\pm$2 sigma in the 3.3 cm image
to better show the extension of emission at this wavelength.}
\label{CB230-14mm}
\end{figure}

Both sources are also detected at 1.4 cm and 3.3 cm as shown in Figure \ref{CB230-14mm}. 
The 3.3 cm detection toward the secondary is marginal ($\sim$3.5$\sigma$); however, the 3.3 cm 
emission appears extended toward the secondary, possibly related to 
the outflow activity in the source. Moreover, at 1.4 cm the primary appears resolved in the 
direction of the outflow axis. The presence of
3.3 cm emission located at the secondary source is an indication 
that there is an outflow driving source, energetic enough to produce
the ionization required to have detectable free-free emission. The flux densities 
measured from the VLA data are given in Table 5 along with Gaussian fitting of the sources.
The CARMA data had lower resolution than the other datasets, $\sim$2\arcsec, but they provided
a shorter wavelength data point. We do not show the 3.4 mm, 4.0 cm, and 6.5 cm data 
data for the sake of brevity, but the flux densities are given in Table 5.

\begin{figure}[!ht]
\begin{center}
\includegraphics[ scale=1.0]{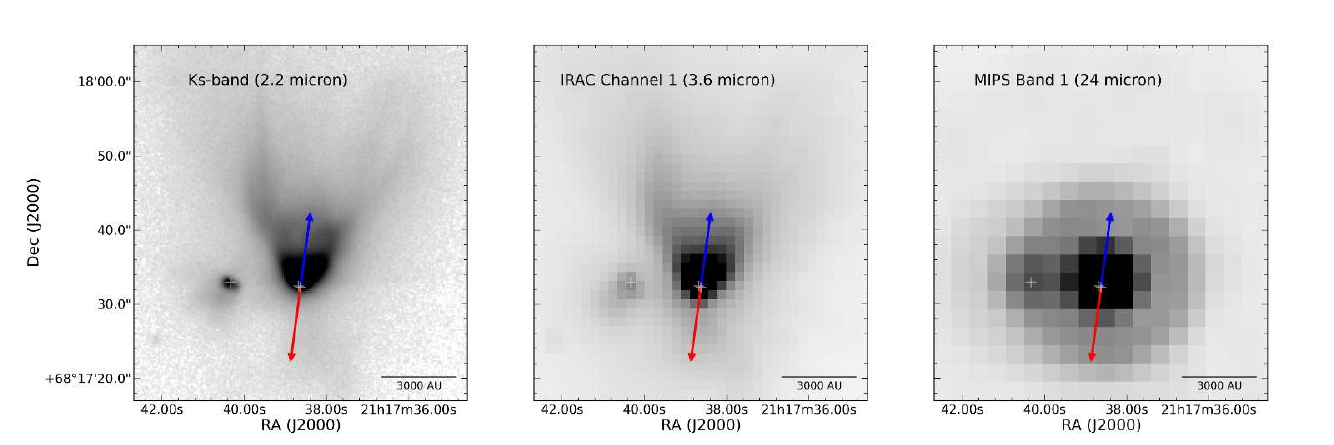}
\end{center}
\caption{Images of CB230 IRS1 and IRS2 at 2.15 \micron\ (left), 4.5 \micron\ (middle), and 24 \micron\ (right). The positions of the two
close binary sources detected by the VLA are the blended crosses near the center of the image and the tertiary source is 10\arcsec\
east. The scattered light emission of the tertiary is clearly seen at the two shorter wavelengths and the source is weakly detected
at 24 \micron, blended with the PSF of the main protostars.}
\label{CB230IR}
\end{figure}

We noted in Section 2 that CB230 IRS1 has an infrared companion (CB230 IRS2) separated from
the main protostar by 10\arcsec\ ($\sim$3000 AU), making CB230 IRS1 a triple system.
Near and mid-infrared images of CB230 IRS1 and IRS2 are shown in Figure \ref{CB230IR}.
We examined the location of the wide tertiary source for emission at all observed wavelengths and
did not detect it; this source is well within the VLA and CARMA primary beams
at 7.3 mm and 3.4 mm (65\arcsec\ and 72\arcsec\ respectively). Therefore, the tertiary source does not
appear to have substantial dust emission or free-free emission, consistent with non-detections
at 1.3 mm and 3 mm \citep{launhardt2001a,launhardt2001b,launhardt2010}; the 3$\sigma$ upper limits
are given in Table 5.

\subsection{L1157-mm}

\begin{figure}[!ht]
\begin{center}
\includegraphics[angle=-90, scale=0.75]{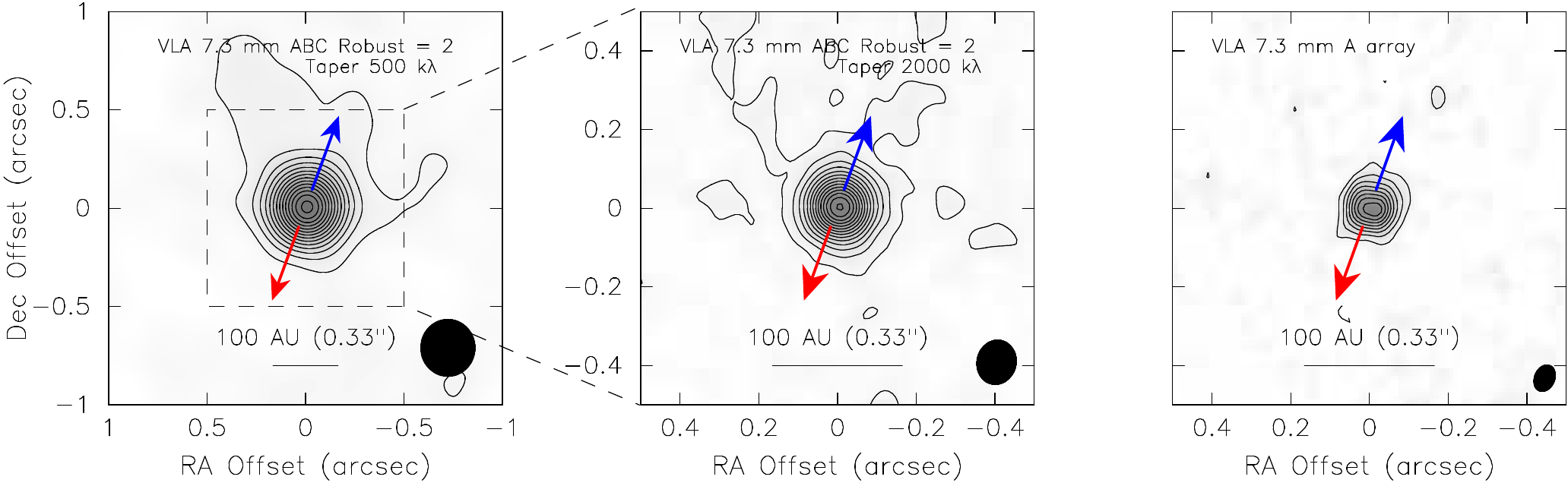}

\end{center}
\caption{L1157-mm images at 7.3 mm. The images are shown with tapering at 500 k$\lambda$ (left), 2000 k$\lambda$ (middle), and 
A-array only untapered, image (right). The tapered images show indications of low-level extended structure
that might be related to warm dust in the outflow cavity as seen a shorter wavelengths \citep{tobin2013,stephens2013}.
There is no evidence of a companion or resolved structure toward the intensity peak in the two tapered images, but
the A-array only image shows a possible indication of resolved structure, consistent with the expected axis of the disk.
The contours in all images start at $\pm$3$\sigma$ and increase in 3$\sigma$ intervals for
all images, where $\sigma$= 38.5 beam$^{-1}$ $\mu$Jy, 24.2 $\mu$Jy beam$^{-1}$, and 18.9 $\mu$Jy beam$^{-1}$ for the 500 k$\lambda$ tapered image,
2000 k$\lambda$ tapered image, and A-array-only image respectively. }
\label{L1157-7mm}
\end{figure}

\begin{figure}[!ht]
\begin{center}
\includegraphics[angle=-90, scale=0.25]{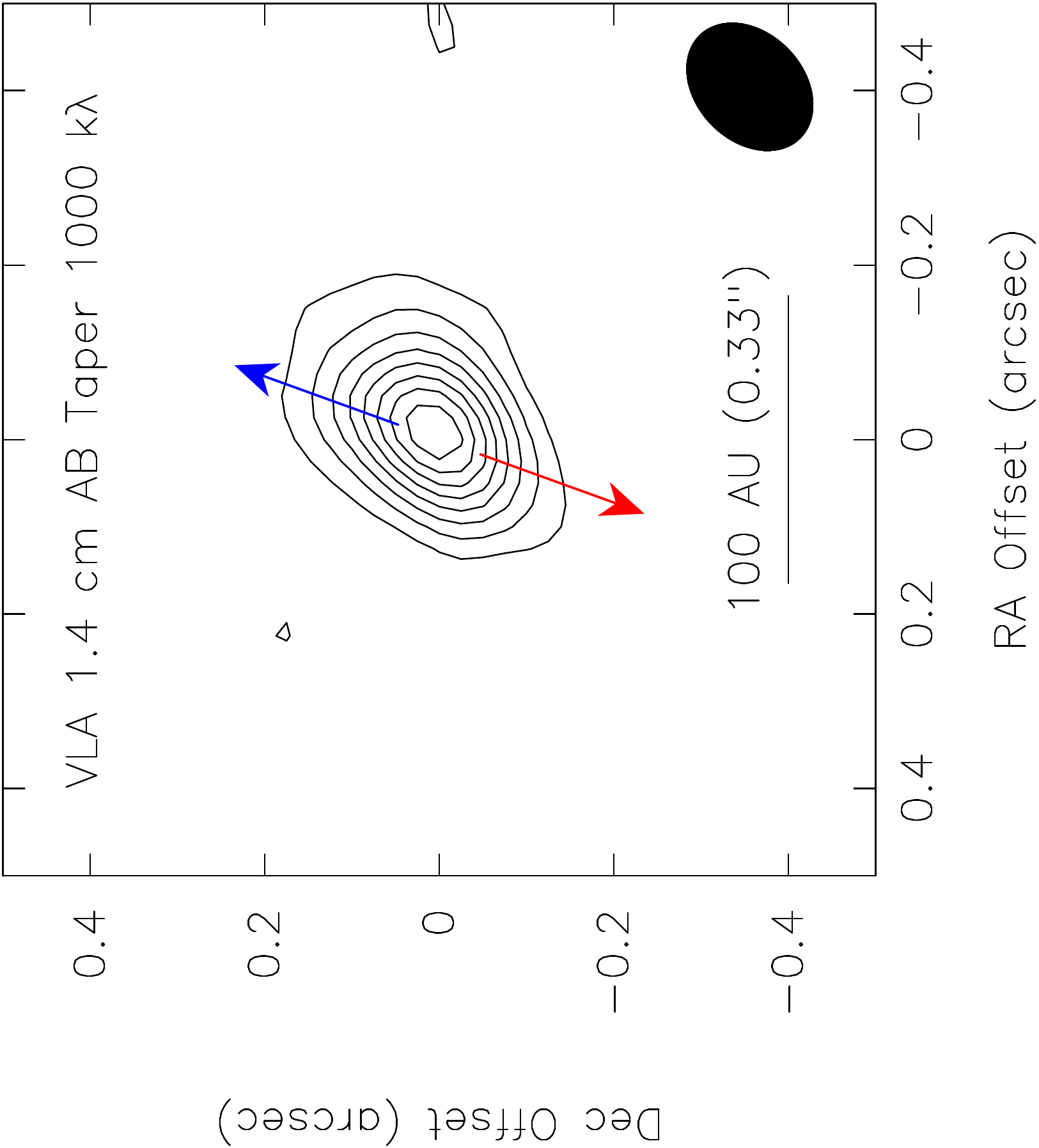}
\includegraphics[angle=-90, scale=0.25]{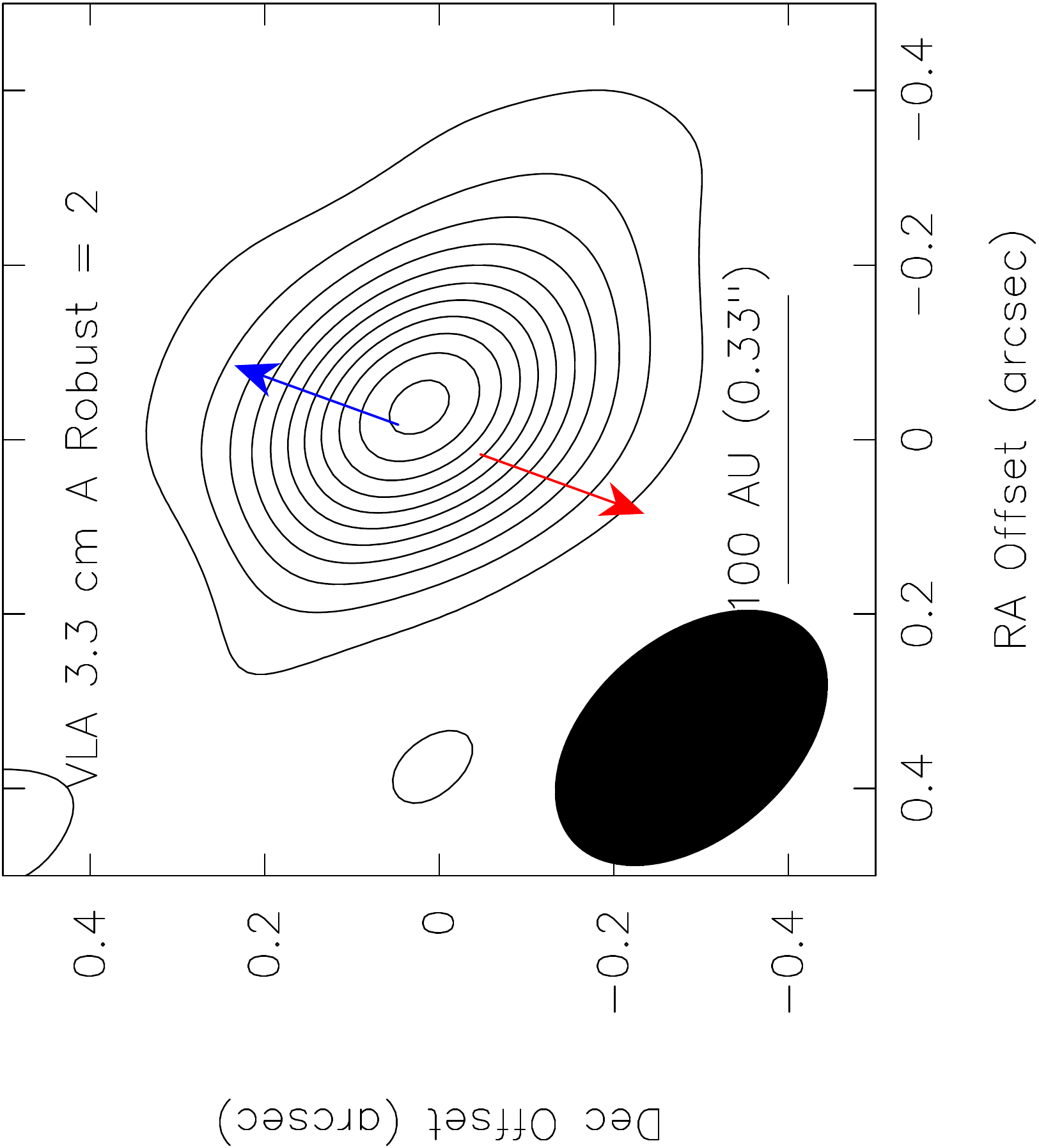}
\end{center}
\caption{L1157-mm images at 1.4 cm (left) and 3.3 cm (right).
The source is marginally resolved at 1.4 cm, but it is compact and does not
show obvious indications of disk-like structure. The contours in all images
start at $\pm$3$\sigma$ and increase in 3$\sigma$ intervals for each image, where
$\sigma_{1.4 cm}$ = 12.1 $\mu$Jy beam$^{-1}$, and $\sigma_{3.3 cm}$ = 6.3 $\mu$Jy beam$^{-1}$. }
\label{L1157-14mm}
\end{figure}

Unlike L1165-SMM1 and CB230 IRS1, L1157-mm appears to be a single source; only one
continuum peak is observed in the 7.3 mm image shown in Figure \ref{L1157-7mm}, as well as the 1.4 cm and 
3.3 cm data in Figure \ref{L1157-14mm}. There is a hint of the source being
marginally resolved perpendicular to the outflow in the A-array-only image 
shown in Figure \ref{L1157-7mm}. Gaussian fitting to the image data (Table 6)
indicates that the source is marginally resolved in the highest resolution 7.3 mm data and the 
deconvolved position angle is perpendicular to the outflow. The marginally extended emission
is suggestive of a small disk (R $<$ 15 AU), but by no means definitive. The lack of
significant resolved structure on small-scales is consistent with the source
having point-like structure at $\sim$0\farcs3 resolution at 1.3 mm and 
3.4 mm \citep{chiang2012}. The 1.4 cm data appear to be
resolved, as evidenced by the peak flux differing from the integrated flux, while the 3.3 cm data are 
consistent with a point source. This could be indicative of the 1.4 cm data
having a marginally extended free-free component as often found by \citet{reipurth2002,reipurth2004}.

\subsection{Millimeter-Centimeter Spectra}

Between 1.3 mm and 3.3 cm, the broad-band spectra of many protostellar objects transition
from one dominated by thermal dust emission to being dominated by free-free emission
from the thermal jets driven by the protostars and disks \citep{anglada1995,anglada1998, shirley2007}.
Emission at 7.3 mm lies in the part of the spectrum where there may be significant 
contributions from both dust and free-free emission. 
In order to interpret the 7.3 mm emission in terms of dust emission alone, we must
subtract the potential contribution from free-free emission.

\begin{figure}[!ht]
\begin{center}
\includegraphics[ scale=0.5]{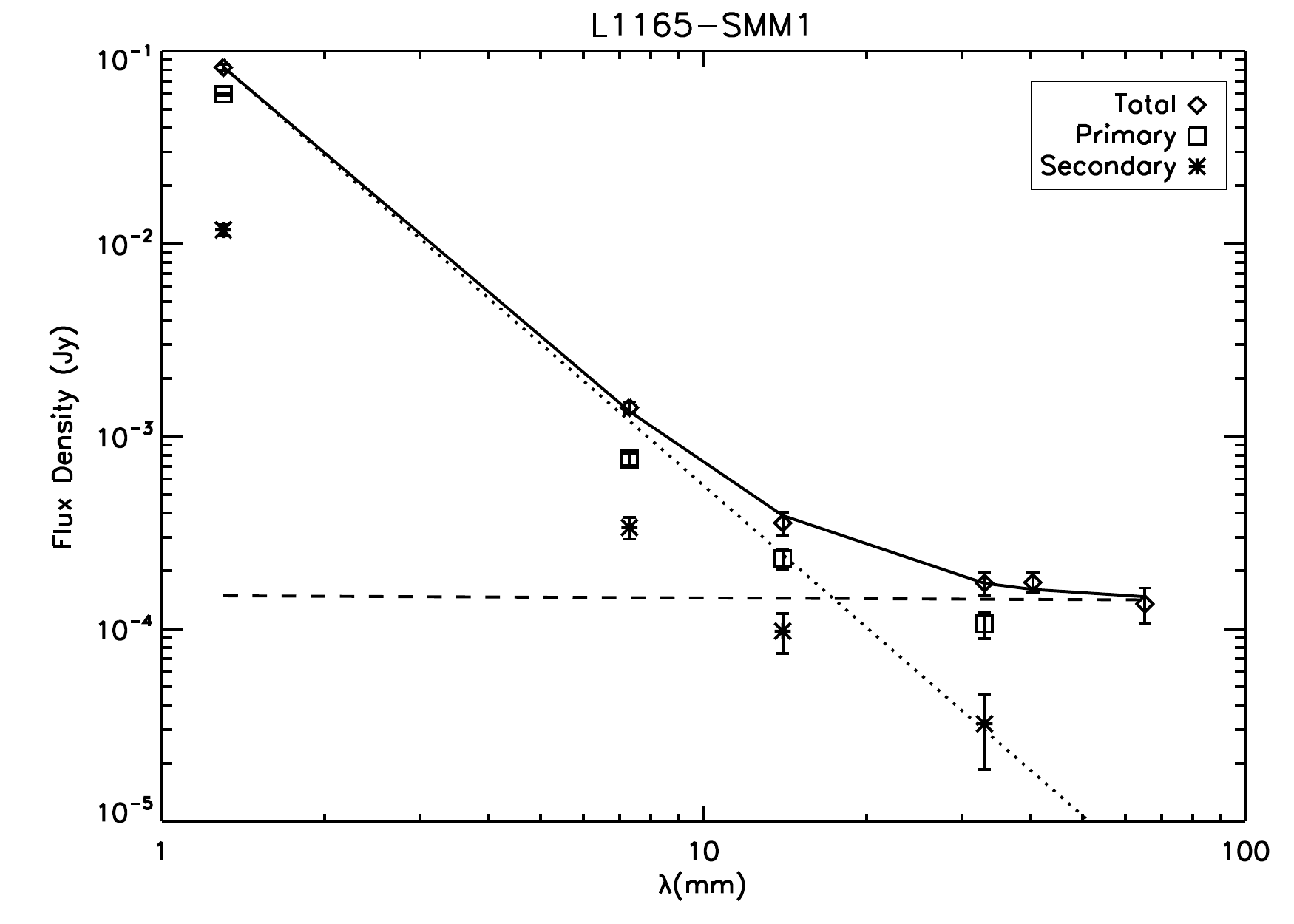}
\includegraphics[ scale=0.5]{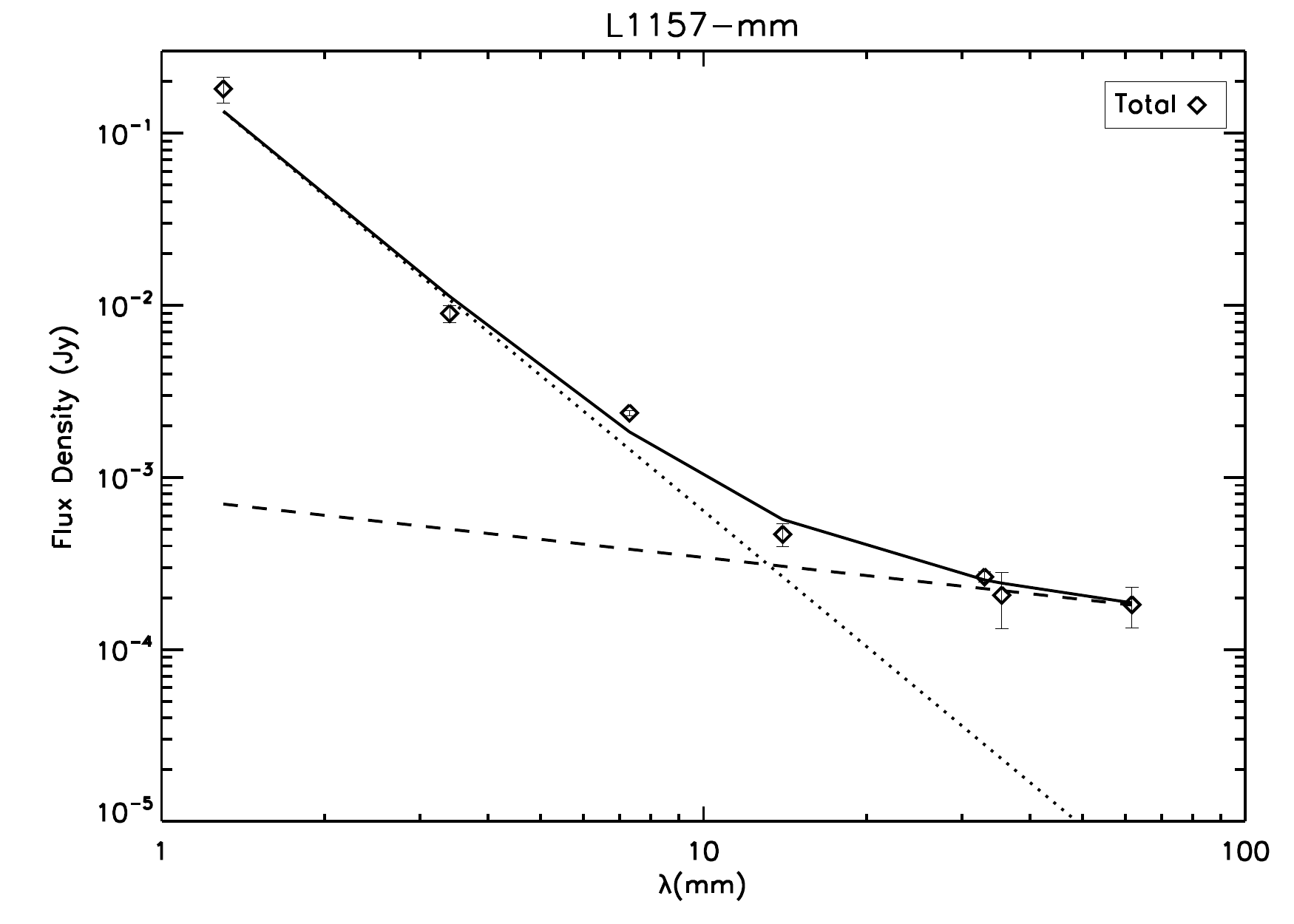}
\includegraphics[ scale=0.5]{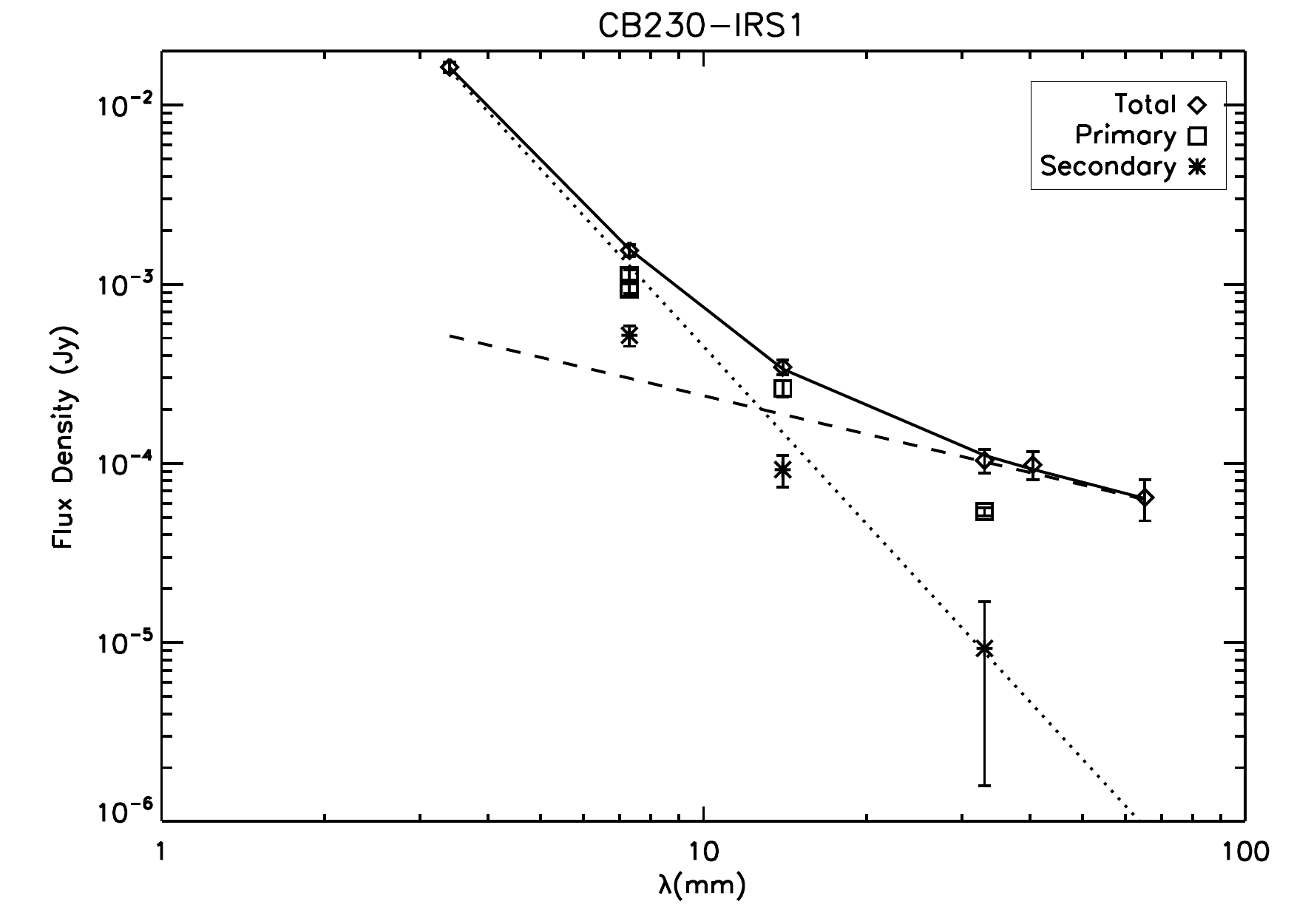}
\end{center}
\caption{Millimeter to centimeter spectra of L1165-SMM1, CB230 IRS1, and L1157-mm. There is a clear free-free (dashed line)
and dust emission (dotted line) component to each spectrum. About 70\% - 90\% of the
emission at 7.3 mm is expected to be from dust continuum (see Section 3.4). The contributions from free-free emission and
dust are about equal at 1.4 cm and the free-free dominates at 3.3 cm.}
\label{SEDs}
\end{figure}

The spectra of the three sources are plotted in Figure \ref{SEDs} from 1.3 mm to 6.5 cm.
The individual flux measurements for the companion sources to L1165-SMM1 and CB230 IRS1
are shown as well as the total flux density. We have attempted to reasonably
match the resolutions of the data from different wavelengths such that we are measuring dust emission 
from the same spatial scales; however, the highest resolution data 
at wavelengths shortward of 7.3 mm is between 0\farcs3 (at best) and $\sim$2\arcsec\ (at worst) for these sources.
Wavelengths longer than 1.4 cm are dominated by unresolved free-free emission, therefore resolution
does not matter as much, except that the multiple sources are not as well-resolved.

We fit the full spectra simultaneously, assuming that the data can be described with two spectral slopes, one representing
the thermal dust emission and the other representing the free-free emission. The fits are made 
using the least-squares fitting routine \textit{mpfit} \citep{markwardt2009}; the 
total fluxes at all the wavelengths plotted in Figure \ref{SEDs}. In the fitting procedure, 
the free-free slope is constrained to be between $\lambda^{-1.0}$ and $\lambda^{1.0}$, while
the thermal slope is assumed to be between $\lambda^{-1.0}$  and $\lambda^{-4.5}$.
Note that this fit is for the total flux density of a source, given that the shorter wavelength 
data do not resolve the proto-binaries. The sources have a range of thermal and free-free spectral slopes
that are given in Table 7. CB230 has a steep spectral slope, while L1165 has a shallow spectral slope;
the spectral slopes in the centimeter are consistent with previous measurements 
toward protostars \citep{shirley2007}. We note that there is significant uncertainty
in the spectral slopes given the degeneracies between the thermal and free-free 
components; moreover, there may be variability in the free-free emission \citep[e.g.][]{shirley2007} 
and that would further add to the uncertainty.

The spectral slope of the free-free emission enables us to subtract its contribution
from the 7.3 mm data to determine the flux density only due to dust emission. This
enables us to estimate the mass of the emitting material with some assumptions. 
Free-free emission at 7.3 mm comprises $\sim$10\% of the total in L1165-SMM1,
$\sim$19\% in CB230 IRS1, and $\sim$18\% in L1157-mm, the values are relative to the flux densities
measured in the 500 k$\lambda$ tapered images. A lingering issue
is that we do not know free-free spectral slope of the companion sources since we can only measure the spectral index
of the total flux. Therefore, we have assumed that the spectral slope of the total flux is applicable to the primary and secondary sources.

\subsection{Mass Estimates}

The masses of the material surrounding each protostar can be estimated by assuming that
the dust emission is isothermal and optically thin, using the formula 
\begin{equation}
M = \frac{D^2 F_{\lambda} }{ \kappa_0\left(\frac{ \lambda }{ 850\mu m }\right)^{-\beta}B_{\lambda}(T_{dust}) };
\end{equation}
$B_{\lambda}$ is the Planck function. We have assumed that $T_{dust}$ = 30 K, $\kappa_{850}$ = 0.035 cm$^{2}$ g$^{-1}$ 
(dust+gas opacity, assuming a dust-to-gas ratio of 100), and D = 300 pc, basing our
assumptions on previous works \citep{beckwith1990, andrews2005, tobin2012}.
The dust opacity spectral index ($\beta$) parameter is estimated from the fit to the spectrum thermal slope (Table 7),
under the assumption that F$_{\lambda}$ $\propto$ $\lambda^{-(2 + \beta)}$. Note
that the dust opacity we have assumed is larger than typically assumed for protostars 
\citep[e.g.][]{ossenkopf1994} and more typical of the Class II sources
\citep[e.g.][]{andrews2005, andrews2009}; however, a larger opacity may be correct given the
indications of grain growth in the protostellar phase \citep{kwon2009, tobin2013}. With this
opacity we can more directly compare our results to those of
more evolved disks, but we may be systematically underestimating the dust masses if the 
dust opacities are indeed lower.

The derived masses are in the range 0.01 to 0.2 $M_{\sun}$, comparable to the
masses found for Class II disks \citep[e.g.][]{andrews2010}. 
The mass estimates for each source and their components are listed in Table 8, along
with the assumed $\beta$. The mass estimates agree
reasonably well at the different wavelengths for all sources;
differences likely reflect a combination of uncertainties in absolute flux calibration, 
variations in dust opacities, sensitivity to different spatial scales, etc. Note that
these mass estimates are temperature dependent and larger characteristic temperatures
would yield lower mass estimates. Moreover, these sources certainly have temperatures gradients
that cannot be accounted for in this simple analysis.

\section{Discussion}

Two protostars out of our sample of three are found to have a previously unknown companion source separated 
by 0\farcs3 ($\sim$100 AU). The presence of these companions raises the questions: how did these companions
form and did they form in situ, or have their separations evolved significantly via dynamic evolution.

\subsection{Constraining the Probable Binary Formation Mechanism}

There are three likely mechanisms for the formation of binary/multiple protostellar systems.
The first is fragmentation induced by the rotating collapse of the infalling envelope
 \citep[e.g.][]{bb1993, boss1995,tohline2002,sterzik2003}. In this scenario, a rotating,  
spherical envelope has a mild azimuthal density perturbation and simulations show that
the density perturbation becomes enhanced during collapse. The rotation causes the formation of a large,
flattened, and asymmetric density structure from which two density peaks form
with $\sim$1000 AU separations. While these simulations are simplistic and have rather 
high initial rotation rates, envelopes
have been observed to be even more asymmetric than assumed 
in these simulations \citep{tobin2010a, launhardt2013}. 
Therefore, this mechanism could form both wide ($\sim$1000 AU) and close ($<$ 500 AU) 
binary systems, depending on the angular
momentum of the infalling gas. The wide binary systems could then further
evolve to smaller separations via dynamical friction, or if the infalling matter had high 
angular momentum, the binary separations might widen rather than shrink \citep{zhao2013}.

The second mechanism is turbulent fragmentation. \citet{offner2010}
found that protostars forming in global molecular cloud simulations
typically had initial separations of a few thousand AU and
fragmentation on the scale of the disk could be suppressed by radiative heating. 
These protostars could then migrate toward
smaller separations via dynamical friction and the accretion of low
angular momentum gas. The dynamical evolution could shrink the separations of systems 
from $\sim$2000 AU to $\sim$200 AU on timescales as short as $\sim$10 kyr. Thus,
even on the short timescale of protostellar collapse, binaries could migrate to small separations
rapidly enough that it would be unclear if they formed in-place or if they had
migrated.

The third formation mechanism is direct fragmentation of the rotationally-supported circumstellar
disk via gravitational instability \citep[e.g.][]{kratter2010, vorobyov2010,zhu2012}.
This is similar to the fragmentation of the rotating, asymmetric envelope, but is happening within the
rotationally-supported circumstellar disks, whereas the flattened envelope was not supported by its rotation.
In this scenario, a large, massive disk ($M_d$ $\sim$ few $\times$ 0.1 $M_{\sun}$)
forms due to the angular momentum of the infalling envelope. The self-gravity 
of the disk enables fragmentation to happen at large radii, forming the binary components in-place
and migration of the fragments could take place within the disk. The fragments would also
all be in the equatorial plane of the system 
and would share a common angular momentum vector. This is also the case for rotational fragmentation
of a common infalling envelope. However, there is no expectation for protostars formed via turbulent
fragmentation to share the same orbital plane/angular momentum vector as their companions. 

We suggest that the likely fragmentation mechanism for CB230 IRS1 and 
L1165-SMM1 is direct fragmentation of their
circumstellar disks. The principle evidence for this is the apparent
 circumbinary disk around the two protostars in L1165-SMM1 and that the secondary sources
in both CB230 IRS1 and L1165-SMM1 being located nearly orthogonal to
the outflow direction. This is expected if they formed in a disk and the outflow is launched
perpendicular to the equatorial plane of the disk. While we did
not detect evidence of a circumbinary disk around CB230 IRS1, its presence or lack thereof cannot be ruled out with
the data in hand. The case for a circumbinary disk around L1165-SMM1 is reasonably strong given that 
the emission is not simply comprised of two blended point sources as shown by our image
decomposition in Figure \ref{L1165-1mm}. Moreover, the total flux is 10 mJy lower
in the robust = -2 image than in the robust = 1 image, indicative of extended structure
being filtered-out at higher resolution. Lastly, there is an indication
of rotation on the scale of the circumbinary disk from the $^{13}$CO ($J=2\rightarrow1$) 
channel maps in Figure \ref{L1165-13CO}. We note that rotationally-supported 
circumbinary disks have been observed toward GG Tau (\citep{guilloteau1999} and 
L1551 NE \citep{takakuwa2012}; therefore,
it is conceivable that the disk around L1165-SMM1 is rotationally supported.
While the masses of material surrounding L1165-SMM1 
and CB230 IRS1 inferred from dust emission are not large, it is important to remember that
much of the mass that was present prior to fragmentation 
has likely already been Incorporated into the protostellar 
objects and we are simply observing the leftovers. 

We concede that even with the evidence presented we cannot
 concretely rule-out envelope rotational fragmentation or turbulent fragmentation with 
subsequent migration, though these processes seem unlikely for the following
reasons. Rotational fragmentation of the envelope is probably 
unlikely since the envelope would have high angular momentum, and this would keep the 
sources from migrating toward smaller separations until the envelope had 
dissipated \citep{zhao2013}. The extended emission observed in single-dish bolometer 
maps shows that the two binary sources are still embedded in 
their natal envelopes \citep{visser2002,launhardt2010}. Turbulent fragmentation is also
probably not likely since migrating companions would not have preferential alignment with
the expected disk plane; however, we cannot absolutely rule-out chance alignment with 
the disk plane for a sample of two.

We note that both CB230 IRS1 and L1165-SMM1 were observed to have large velocity gradients ($\sim$ 11 \kmspc) perpendicular
to the outflow directions in their respective envelopes, relative to the full samples presented in \citet{tobin2011,chen2007}.
L1157-mm has a smaller velocity gradient (3.5 - 6.2 \kmspc) and is found to be single. Under the assumption
that the velocity gradients trace pure rotation and the central protostellar masses were 0.5 $M_{\sun}$, 
the centrifugal radii of the material at 2000 AU, were $\sim$100 AU, $\sim$150 AU, and $\sim$50 AU for
L1165-SMM1, CB230 IRS1, and L1157-mm respectively \citep{tobin2012}. These values agree reasonably well with the
scale of the binaries observed in L1165-SMM1 and CB230 IRS1 as well as the size of the circumbinary disk structure
in L1165-SMM1. Moreover, another close binary candidate HH211 \citep{lee2009} was also found to have a large velocity
gradient normal to the outflow \citep{tanner2011, tobin2011}.  While we questioned the interpretation
of velocity gradients as pure rotation in \citet{tobin2011} and \citet{tobin2012}, the
coincidence of binary/multiple systems where there are large velocity gradients is suggestive
of a relationship between the two, but better statistics are needed.

Our results taken with previous studies increase the number of mm/cm-wave 
proto-binary systems with separations $<$ 150 AU to 7, the others 
being L1551 NE, L1551 IRS5, IRAS 16293-2422A, HH211, VLA1623; L1527 had been previously
suggested to be a 25 AU binary \citep{loinard2002}, but recent VLA follow-up 
has ruled this out (Melis et al. 2013 in prep.). \citet{maury2010}
had suggested that there was a lack of binary systems within the specific separation range
of 150 AU $<$ R $<$ 550 AU. While this range is quite specific, there are definitely
systems with separations smaller and our detections are near the lower end of this limit. \citet{zhao2013}
had suggested that a prevalence of binary systems at radii $\la$100 AU might result from
accreting material that has been magnetically braked and could explain the results of \citet{maury2010}.
On the other hand, a large archival study by \citet{chen2013} finds systems with a variety
of separations, and within the region where \citet{maury2010} suggested a deficit. Our results
combined with the other studies indicate that the result from \citet{maury2010} may have
simply resulted from small number statistics.

We note that there is uncertainty in the distance to the sources in this work; however, all distance work
thus far \citep{kun1998,kun2009,straizys1992} does not point to the sources being a factor of two closer, but
the sources could be a factor of $\sim$1.75 more distant. This uncertainty does not strongly affect 
any conclusions in this work since the size of resolved structure increases linearly with distance and 
masses would increase with the square of the distance. Thus, the separation of the companions is
at most 200 AU, still within an expected scale for circumstellar disks to form.

\subsection{Disk Sizes}

The VLA and CARMA observations of L1165-SMM1 have revealed a possible $\sim$200 AU diameter
circumbinary disk with a total mass of $\sim$ 0.03 $M_{\sun}$; the mass
of just the circumbinary component from image decomposition is $\sim$ 0.007 $M_{\sun}$. 
The higher resolution VLA observations
also show an apparent $\sim$40 AU diameter disk-like structure surrounding the primary source. 
These structures are strong, circumstantial evidence for the presence of a rotationally-supported
disk around this Class 0/I protostar. There is evidence of rotation in the $^{13}$CO data shown in 
Figure \ref{L1165-13CO}, as well as \hcop\ ($J=1\rightarrow0$) observations that
detect high-velocity ($\sim$3 \kms) blue and red-shifted emission located in 
the plane of the disk \citep{tobin2011,tobin2012}. In CB230 IRS1, we observe the presence of a 
resolved structure $\sim$40 AU in diameter surrounding the secondary
source, but the primary appears unresolved, indicating a circumprimary disk
size of less than 30 AU in diameter. We lack complementary observations at shorter wavelengths that might
detect a circumbinary disk, if present. L1157-mm does not have strong indications of a disk-like structure at 7.3 mm, 
consistent with its lack of resolved disk structure at 1.3 mm and 3.4 mm \citep{chiang2012}; therefore,
any disk in L1157 likely has R $<$ 15 AU.

A source of uncertainty in the lack of disk detections is whether or not the 7.3 mm data
are simply `missing' the larger disks due to insufficient sensitivity. 
Studies of Class II disks show that the 8 mm dust continuum
emission from T-Tauri disks can be more spatially compact than the emission 
at shorter wavelengths, indicating that the longest wavelengths are only tracing the innermost regions 
of the disks where the grains have grown to larger sizes \citep{perez2012}. This has been interpreted
to be consistent with radial drift of large dust grains toward small radii \citep{weiden1977}. Models including
radial drift and grain growth during protostellar collapse and disk formation demonstrate that this process 
could be happening in the protostellar phase as well \citep{birnstiel2010}. Moreover, evidence for grain growth
is also seen in the envelopes before the material even reaches the disk \citep{kwon2009}.

For L1165-SMM1, the 7.3 mm emission observed on similar spatial scales as
the 1.3 mm emission is not as extended, but this may simply be due to our sensitivity limits.
The Class 0 protostar L1527 harbors a $\sim$125 AU diameter rotationally supported disk \citep{tobin2012}
and was found to have a 40 AU diameter disk-like structure detected at 7 mm with the old 
VLA system \citep{loinard2002}. Follow-up VLA observations of L1527 at 7 mm detect the disk 
out to $\sim$60 AU (Melis et al. 2013 in prep.).
Thus, observations at $\lambda$ = 7 mm may systematically underestimate the sizes of protostellar disks. The resolved
structures are best regarded as lower limits to the disk size and indicate that observations at shorter
wavelengths may detect yet larger structures, as was the case for L1527 and L1165-SMM1. Given that L1157-mm lacks
resolved structure at shorter wavelengths and in the VLA data we can confidently conclude that its disk is
small. On the other hand, CB230 IRS1 may have a larger circumbinary disk that we do not detect at
7.3 mm, given its similarities to L1165-SMM1.

Since our sample was quite small and not all are bonafide Class 0 protostars, 
we cannot draw broad conclusions about the sizes of disks in Class 0
sources. However, given the youth of L1165-SMM1, it appears to have formed a large disk by the
time it transitioned to the Class I phase. The compact circumprimary disk likely formed at the 
same time as the larger circumbinary disk. Moreover, if the companion source formed via disk fragmentation,
the disk would have had to be more massive in the past given the currently low mass of
the circumbinary structure.

 Recently a few other Class 0 protostars
have been found to harbor disks: L1527 in Taurus \citep{tobin2010b, tobin2012} having a confirmed
Keplerian disk, HH211 having disk-like structure and possible rotation \citep{lee2009}, VLA 1623 A
having resolved structure and possible Keplerian rotation \citep{murillo2013}, finally IRAS 16293-2422B
appears to have a face-on, optically thick disk with R $\sim$ 25 AU \citep{zapata2013}. \citet{maury2010} had a sample of 5 sources
that they concluded did not show evidence of disks or binaries, but we now know that L1527
does have a disk, so their disk-less sample is reduced to four. Combining the results from these studies,
4 systems have evidence of large (R $>$ 100 AU) disks having formed during the Class 0 phase
and 6 systems do not have evidence of large disks. Therefore, the rather strong conclusion by \citet{maury2010}
that disk formation is suppressed by magnetic fields during the Class 0 phase does not seem to be supported by
more recent results. Rather it appears that there may be a wide diversity in the sizes of Class 0 disks, likely
reflecting a variety of initial formation conditions. Note that we have left CB230 IRS1 out of these numbers since
we do not know if there is an undetected circumbinary disk present.

\subsection{A Disrupted Triple System in CB230?}

We noted in Section 3.2 that there is an infrared-detected tertiary component (CB230 IRS2) separated 
from the main protostar (CB230 IRS1) by $\sim$10\arcsec\ (3000 AU). This source was not 
detected at millimeter or centimeter wavelengths indicating
that the dust mass from this source must be $<$ 0.005 $M_{\sun}$ (3$\sigma$ upper limit using
assumptions from Section 3.5) and that any free-free emission from this source
must be very weak, indicating that the source is not driving a strong
thermal jet. This leads us to examine its relationship with the protostellar system
more closely. The tertiary is a highly-reddened source, not detected in the optical,
with strong infrared excess emission at 24 \micron\ \citep{massi2008}; it is resolved in 
the near-infrared and there is some diffuse scattered light emission around the source (Figure \ref{CB230IR}).
Moreover, \citet{launhardt2001b}
found indications of two outflows in $^{13}$CO observations, while not detecting
the tertiary in continuum \citep{launhardt2001b,launhardt2010}. 
Moreover, recent $^{12}$CO ($J=2\rightarrow1$) also sees
strong evidence for an outflow orthogonal to the outflow 
from the main source (Segura-Cox et al. in prep.).

The near to mid-infrared emission can be explained by a combination of direct stellar
light, scattered light, and warm dust emission from an inner disk. Infrared excess emission
from the inner disk can be produced with disk masses $<$ 0.001 $M_{\sun}$ \citep{robitaille2007,espaillat2010}.
Thus, all evidence points to the tertiary being young and in close association with CB230 IRS1.
Unlike other protobinaries at large separations (e.g. NGC 1333 IRAS 2, 
L1448 IRS3, BHR71) it does not show the typical signs of extreme youth like IRS1. There is little evidence of it  
accreting from the large infalling envelope; there is not increased linewidth
or velocity structure related to the wide companion in \nthp\ or ammonia \citep{chen2007,tobin2011} and
the far-infrared emission is not extended toward its location \citep{launhardt2013}.

Given its close association with the CB230 IRS1 and its lack of Class 0/I protostellar properties,
we suggest that the tertiary source may have formed in closer proximity to the compact binary system, perhaps
in a disk and was ejected in a three-body interaction \citep{reipurth2000b,reipurth2001}.
Simulations of disk fragmentation in protostellar
systems have produced such ejections of low-mass members via interactions with the primary source and 
other disk fragments \citep{basu2012}. Such an interaction could leave the tertiary
with very little circumstellar material \citep{reipurth2000b}, 
consistent with the lack of detectable millimeter-wave emission.
If the tertiary source was kicked out with a velocity of $\sim$1 \kms\ \citep{basu2012}
it could have reached its present position in $\sim$15 kyr, well within the expected $\sim$500 kyr evolutionary
timescales of the system as a whole \citep{evans2009}. Furthermore, the location of the tertiary is approximately
orthogonal to the outflow, in the same plane as the compact secondary 
and expected disk plane. However, the outflow indicates
that the current angular momentum vector of the tertiary may be perpendicular to that of the primary.
This hypothesis of the tertiary source being an ejected member could be tested with high-resolution
near-infrared spectra of the primary and secondary sources, enabling the relative radial velocities of
the sources to be examined. However, it is uncertain if sufficiently accurate velocities could be 
measured toward these sources, given that the near-infrared emission is dominated scattered light and 
\citet{covey2006} was only able to obtain precisions of $\sim$2 \kms\ for Class I protostars.

\section{Conclusions}
We have presented a VLA and CARMA study of three Class 0/I protostellar systems 
in the Cepheus clouds: L1157-mm, L1165-SMM1, and CB230 IRS1. 
The VLA observations were taken in A, B, and C configurations at 7.3 mm, 1.4 cm, 
3.3 cm, 4.0 cm, and 6.5 cm providing the
highest resolution view of these systems at any wavelength. The CARMA observations in B and C-arrays
toward L1165-SMM1 have also provided the highest resolution view of the system at 1.3 mm. 
The results presented have enabled us to expand our knowledge of the disk and binary
star formation process by probing scales $<$ 100 AU and a 3$\sigma$ 
mass sensitivity 0.005 $M_{\sun}$, an order of magnitude more sensitive than 
the typical mass of a T-Tauri disk \citep{andrews2009}.  
Moreover, the sizes of the resolved structures from the VLA and CARMA
observations provide lower limits on disk sizes during the early stages of protostellar evolution.

We find that 2 (L1165-SMM1 and CB230 IRS1) out of 3 observed systems have an apparent companion separated by 0\farcs3 ($\sim$100 AU). 
The companions are well-detected and resolved from the primary at 7.3 mm and 1.4 cm; they are weakly detected
at 3.3 cm and somewhat blended with the primary components. The 1.3 mm imaging of L1165-SMM1 find an extended disk
structure surrounding the two components and possibly two outflows. Our results point
to a disk fragmentation origin of L1165-SMM1 and CB230 IRS1, given their close proximity and alignment
with the expected orbital plane of the disk in these systems; this interpretation is stronger
for L1165-SMM1 given the extended disk centered on the primary protostar as observed at 1.3 mm.
CB230 IRS1 does have a third member of its system, detected in the near and mid-infrared and its direction
from the primary source is perpendicular to the outflow; we do not detect emission from this source between 3.4 mm and 6.5 cm. 
We suggest that it may have been ejected via a three-body interaction.

Despite the high sensitivity of our VLA observations, we did not detect direct evidence of 
large ($\sim$100 AU) disk structures surrounding any of the sources at 7.3 mm. The extended 7.3 mm emission
toward L1165-SMM1 is not obviously disk-like and the CARMA 1.3 mm data gave a more firm detection of a
probable circumbinary disk. The primary source
in L1165-SMM1 does have resolved structure, consistent with a small disk ($\sim$40 AU in diameter); 
the secondary source in L1165-SMM1 appears unresolved. The primary source in CB230 IRS1 appears unresolved,
and the secondary source appears to have a resolved structure $\sim$40 AU in diameter. Finally,
L1157-mm does not show strong evidence for a resolved disk, indicating that the disk is at most 15 AU in radius.

We would like to thank the anonymous referee for constructive comments
that improved the manuscript. We would also like to thank S. Offner for useful discussions
regarding the results and C. Brogan for discussions regarding the data reduction.
J. Tobin acknowledges support provided by NASA through Hubble Fellowship 
grant \#HST-HF-51300.01-A awarded by the Space Telescope Science Institute, which is 
operated by the Association of Universities for Research in Astronomy, 
Inc., for NASA, under contract NAS 5-26555. H.-F. C. acknowledges
support from the National Aeronautics and Space
Administration through the NASA Astrobiology Institute under
Cooperative Agreement No. NNA09DA77A issued through the Office of
Space Science. L.W.L. and H.-F. C. acknowledge support from the Laboratory for Astronomical 
Imaging at the University of Illinois and the NSF under grant AST-07-09206.
P. D. acknowledges a grant from PAPIIT-UNAM.
L. L. acknowledges the support of DGAPA, UNAM, CONACyT (M\'exico).
T. B. acknowledges support from NASA Origins grant NNX09AB89G.
Support for CARMA construction was derived from the states of Illinois, California, and Maryland, 
the James S. McDonnell Foundation, the Gordon and Betty Moore Foundation, the Kenneth T. and 
Eileen L. Norris Foundation, the University of Chicago, the Associates of the California 
Institute of Technology, and the National Science Foundation. Ongoing CARMA development 
and operations are supported by the National Science Foundation under a cooperative 
agreement, and by the CARMA partner universities. The National Radio Astronomy 
Observatory is a facility of the National Science Foundation 
operated under cooperative agreement by Associated Universities, Inc.

{\it Facilities:}  \facility{VLA}, \facility{CARMA} \facility{Hiltner}, \facility{Spitzer}

\appendix
\section{L1165-SMM1 Photometry}

We have collected relevant photometry for L1165-SMM1 and constructed the spectral energy distribution
between 2.15 \micron\ and 850 \micron. Note that we do not include the interferometer data in this
analysis since they may resolve-out emission from the envelope; moreover, the flux densities are low
enough that they do not significantly contribute to the bolometric luminosity or temperature calculations.
The 850 \micron\ data for L1165-SMM1 are from \citet{visser2002}, the 25 \micron\, 60 \micron\, and 100 \micron\
points are from IRAS \citep{iras}, and the 12 \micron\ and 22 \micron\ data are from the WISE survey 
\citep{wise}; the additional data are described in the following subsections.

\subsection{Near-infrared Data}

The Ks-band data were taken at the MDM observatory on 2008 June 23, the data were reduced using standard methods
for near-infrared imaging, details of the reduction and observational methods
can be found in \citet{tobin2008} and \citet{tobin2010a}.

\subsection{Spitzer Data}
The \textit{Spitzer} observations and reduction for L1165-SMM1 were presented in \citet{tobin2010a}.
The IRAC data cover 4 near/mid-infrared bands: 3.6 \micron, 4.5 \micron, 5.8 \micron, and 8.0 \micron. 
We performed photometry on the IRAC data in 10000 AU apertures (33.3\arcsec) and applied the extended 
source aperture corrections\footnote{http://irsa.ipac.caltech.edu/data/SPITZER/docs/irac/iracinstrumenthandbook/30/}.
The MIPS data at 24 \micron\ and 70 \micron\ are taken from the Cores-to-Disks Legacy program data delivery and photometry
 were taken in aperture radii outside the first airy ring and the requisite aperture corrections were 
applied\footnote{http://irsa.ipac.caltech.edu/data/SPITZER/docs/mips/mipsinstrumenthandbook/home/}.

\subsection{Herschel Data}
\begin{figure}[!ht]
\begin{center}
\includegraphics[ scale=1.25]{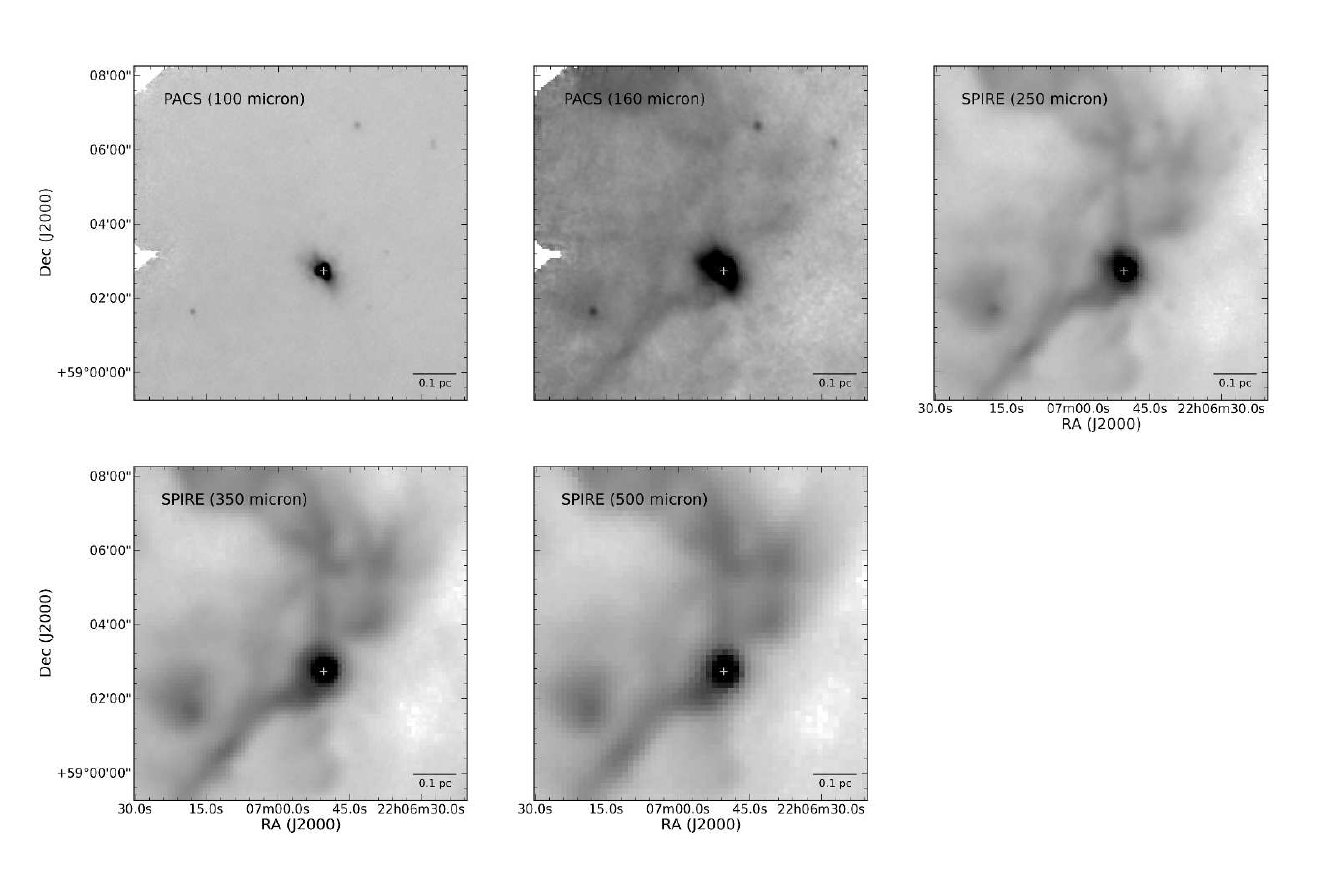}
\end{center}
\caption{L1165 \textit{Herschel} maps at 100 \micron, 160 \micron, 250 \micron, 350 \micron, and 500 \micron. L1165-SMM1 is
the bright source marked with a white cross in each map.}
\label{L1165-Herschel}
\end{figure}

\begin{figure}[!ht]
\begin{center}
\includegraphics[ scale=0.5]{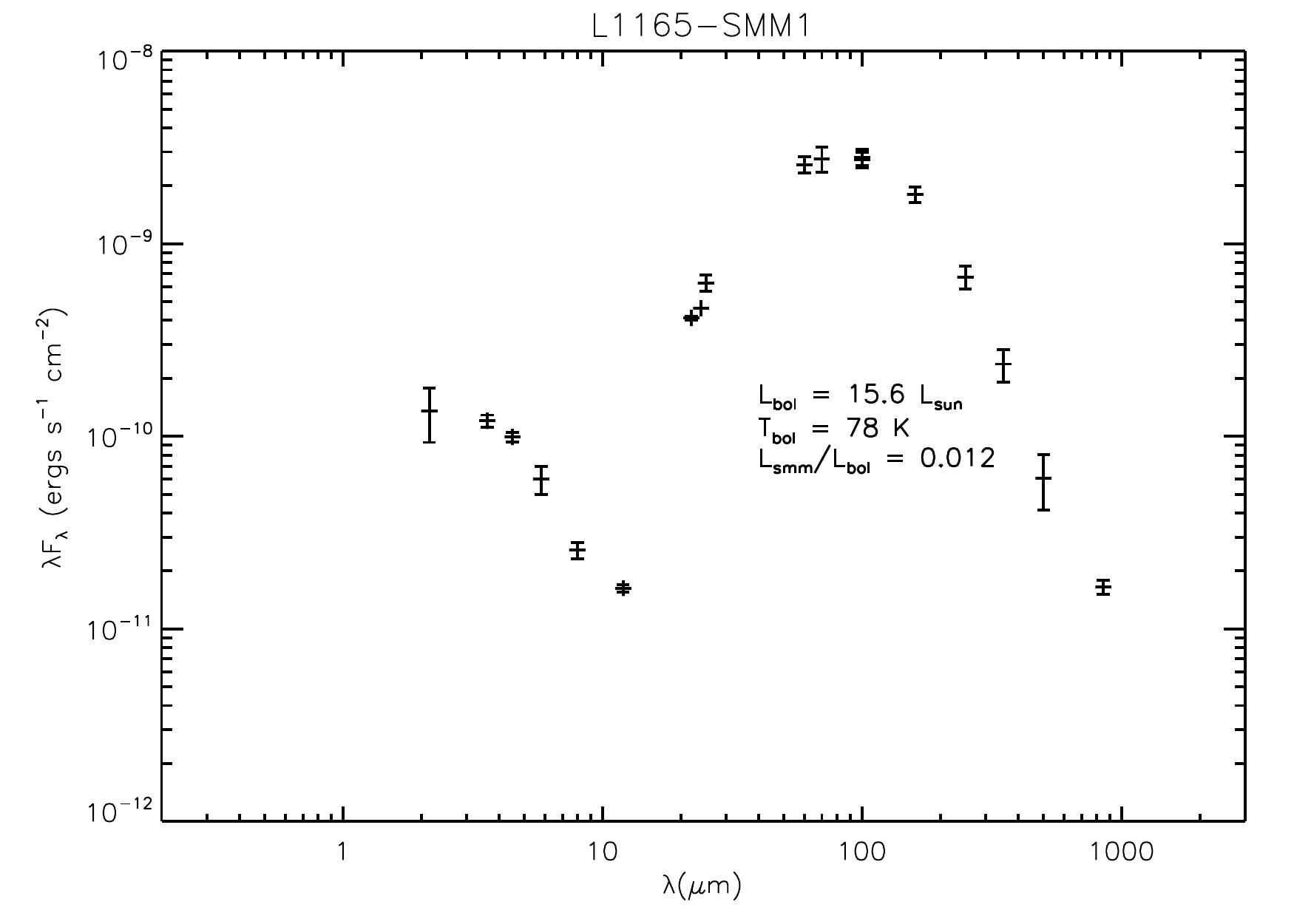}
\includegraphics[ scale=0.5]{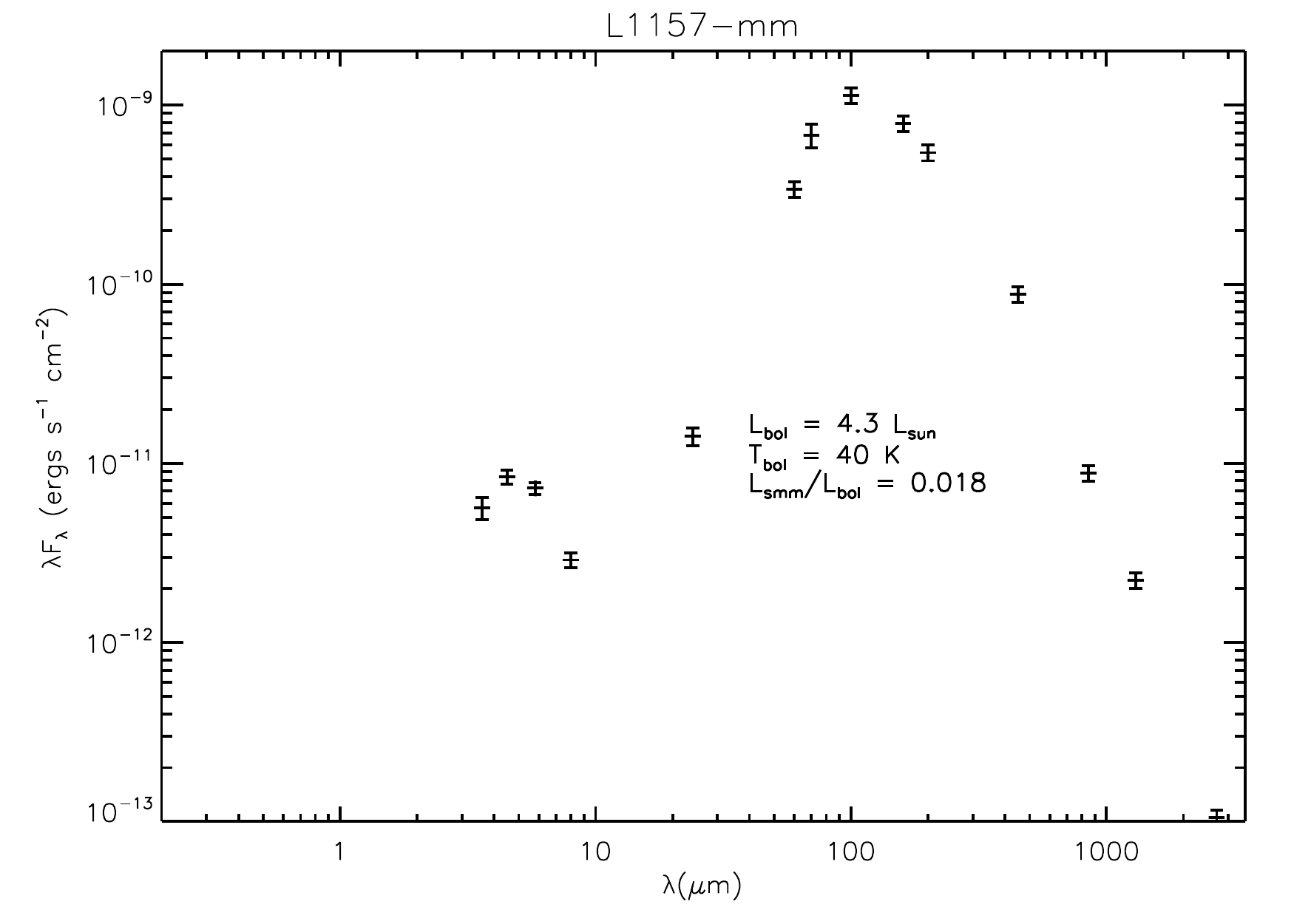}
\end{center}
\caption{SED plots of L1165 and L1157 including all photometry used in the derivation of T$_{bol}$ and L$_{bol}$.}
\label{tbolseds}
\end{figure}

The L1165 dark cloud was observed with the \textit{Herschel} PACS and SPIRE photometers on 2011 May 05 and 2011 July 11; 
observation ids are 1342223967, 1342223968, and 1342219969. The PACS and SPIRE maps were generated using the Scanamorphos software
version 20 with the galactic option. The SPIRE level 1 data were processed directly by Scanamorphos and
the PACS data were converted to Scanamorphos format using the \textit{convertL1Scanam} procedure in HIPE version 10. These
reductions made use of PACS calibration version 45 and SPIRE calibration version 10.1. The maps of the five \textit{Herschel}
bands are shown in Figure \ref{L1165-Herschel}. The PACS photometry were taken in 12\arcsec\ 
apertures at 70 \micron\ and 100 \micron\ and a 22\arcsec\ aperture at 160 \micron\ as
recommended by the PACS calibration documentation\footnote{http://herschel.esac.esa.int/twiki/bin/view/Public/PacsCalibrationWeb}.
SPIRE photometry were taken in 22\arcsec, 30\arcsec, and 42\arcsec\ radii in the 250 \micron, 350 \micron, and 500 \micron\
bands respectively with the requisite aperture corrections for extended sources and color corrections 
applied\footnote{'http://herschel.esac.esa.int/hcss-doc-9.0/load/spire\textunderscore drg/html/ch05s07.html'}.
The uncertainty in the absolute flux scale for SPIRE and PACS is estimated to be $\sim$5\%. The SED plot for L1165-SMM1
is shown in Figure \ref{tbolseds}.

\section{L1157 Photometry}

We have also collected the relevant photometry to construct a full SED of L1157. The \textit{Spitzer} IRAC
and MIPS data are from \citet{kirk2009} and the 450 \micron\ and 850 \micron\ data are from \citet{young2006}.
We also include 60 \micron, 100 \micron, 160 \micron\, and 200 \micron\ data from the \textit{Infrared Space Observatory} \citep{froebrich2005}.
The SED plots for L1157-mm is shown in Figure \ref{tbolseds}.

\section{Bolometric Temperature}

The bolometric temperature is a standard observational diagnostic for young stellar objects to determine their observational class, complementing
the spectral index and submillimeter luminosity diagnostics. The bolometric temperature is defined as the temperature of a blackbody
having the same average frequency. \citet{myers1993} calculate the average frequency
\begin{equation}
\langle \nu \rangle = \frac{\int \nu S_{\nu} d\nu}{\int S_{\nu} d\nu}
\end{equation}
where $S_{\nu}$ is the observed flux density at a given frequency and then the bolometric temperature
\begin{equation}
T_{bol} = 1.25 \times \langle \nu \rangle\ \mathrm{K}\ \mathrm{Hz^{-1}}.
\end{equation}
\citet{chen1995} then defined approximate Class boundaries at $T_{bol}$ $\sim$ 70 K for Class 0 to Class I,
 $T_{bol}$ $\sim$ 650 K for Class I to Class II, and $T_{bol}$ $\sim$ 2800 K for Class II to Class III.

To calculate the integrals in equation C1, we use the trapezoidal integration routine 
\textit{tsum}, found in the IDL Astronomy Library \citep{landsman1993}. 
We calculate $T_{bol}$ = 40 K, $L_{bol}$ = 4.3 $L_{\sun}$, and $L_{submm}$/$L_{bol}$ = 0.018 for L1157; for L1165,
$T_{bol}$ = 78 K, $L_{bol}$=15.6 $L_{\sun}$, and $L_{submm}$/$L_{bol}$ = 0.012.

\begin{small}
\bibliographystyle{apj}
\bibliography{ms}
\end{small}

\clearpage

\begin{deluxetable}{lllllllllll}
\tablewidth{0pt}
\rotate
\tabletypesize{\scriptsize}
\tablecaption{Source List}
\tablehead{
  \colhead{Source}  & \colhead{RA} & \colhead{Dec}      & \colhead{Distance}    &  \colhead{Mass$_{8\mu m}$\tablenotemark{\dagger}} & \colhead{Mass$_{submm}$\tablenotemark{*}}   & \colhead{L$_{bol}$}   & \colhead{L$_{submm}$/L$_{bol}$}   & \colhead{T$_{bol}$}  & \colhead{Outflow PA} & \colhead{References}\\
                    & \colhead{(J2000)} &  \colhead{(J2000)}     & \colhead{(pc)}          &  \colhead{($M_{\sun}$)}     & \colhead{($M_{\sun}$)}                      & \colhead{($L_{\sun}$)}&  & \colhead{(K)}        & \colhead{(\degr)} & \colhead{(Distance, M$_{ref}$,}  \\
                    &                   &     &                                           &  \colhead{(r$<$0.05pc)}     &                                    &                                   &                            &              & & \colhead{Outflow PA)}
}
\startdata

CB230 IRS1         & 21:17:38.56 & +68:17:33.3 & 300  & 1.1  & 1.1         & 5.85  & 0.037  & 230      & 350 & 5, 6, 8\\
L1157-mm           & 20:39:06.25 & +68:02:15.9 & 300  & 2.6  & 2.2         & 4.3  & 0.018  & 40      & 150 & 5, 4,  7\\
L1165-SMM1         & 22:06:50.46 & +59:02:45.9 & 300  & 1.1  & 0.32        & 15.6 & 0.012  & 78      & 225 & 5, 2, 2\\

\enddata
\tablecomments{Positions reflect the coordinates of the brightest source in the 7.3 mm
 continuum continuum position for protostars observed with the VLA. L$_{bol}$, T$_{bol}$, and L$_{submm}$/L$_{bol}$ for L1165 and L1157 
were recalculated in our work based on compiled photometry from the literature and archival data. L$_{bol}$ for CB230 is rescaled to 
a distance of 300 pc from the 400 pc used in \citet{launhardt2013} where they give L$_{bol}$ = 10.4 L$_{\sun}$.
The Outflow position axes (PA) are not well constrained since the outflows are known to precess they can have fairly large
angular width; a conservative estimate of uncertainty would be $\pm$10\degr. 
References: (1) \citet{froebrich2005}, (2) \citet{visser2002},
 (4) \citet{shirley2000}, (5) \citet{kirk2009}, (6) \citet{kauffmann2008},
(7) \citet{gueth1996}, (8) \citet{launhardt2001a}.} 
\tablenotetext{*}{Mass was computed with sub/millimeter bolometer data assuming an isothermal temperature.}
\tablenotetext{\dagger}{Mass measured using the envelope extinction of the 
8 \micron\ background emission in \citet{tobin2010a}.}
\end{deluxetable}

\begin{deluxetable}{lllllll}
\tablewidth{0pt}

\tabletypesize{\tiny}
\tablecaption{VLA Observation Log}
\tablehead{
  \colhead{Sources} &\colhead{Config.}  & \colhead{Date} & \colhead{Track Length} & \colhead{Central Frequency(s)} & \colhead{Gain Calibrator/Flux Density} & \colhead{Flux Calibrator}\\
                    &                   & \colhead{(UT)} &  \colhead{(hrs)}       & \colhead{(GHz)}                  &  \colhead{(Name, Jy)}                  &\\
}
\startdata
L1157                                       & C-array & 26 Feb 1996 & 1.25 &  4.86, 8.46         &  2021+614, 3.1, 2.9   &  3C48      \\
L1165, L1157, CB230\tablenotemark{\dagger}  & C-array & 12 Feb 2012 & 4   &  41 &  J2022+6136, 1.6   &  3C48\tablenotemark{*}      \\
L1165, L1157, CB230\tablenotemark{a}        & C-array & 19 Feb 2012 & 1.5 &  41 &  J2022+6136, 1.6   &  3C48      \\
L1165, L1157, CB230                         & C-array & 12 Mar 2012 & 1.5 &  41 &  J2022+6136, 1.35   &  3C48      \\
\\
CB230\tablenotemark{b}                      & B-array & 04 Jun 2012 & 3 &  9, 21, 41 &  J2009+7229, 0.75, 0.65, 0.74   &  3C48      \\
L1157                                       & B-array & 07 Jun 2012 & 3 &  9, 21, 41 &  J2006+6424, 0.7, 0.56, 0.51   &  3C48      \\
L1165                                       & B-array & 11 Jun 2012 & 3 &  9, 21, 41 &  J2022+6136, 3.0, 1.9, 1.1   &  3C48      \\
\\
L1157                                       & A-array & 07 Oct 2012 & 1 &  9         &  J2006+6424, 0.47   &  3C286      \\
CB230                                       & A-array & 11 Oct 2012 & 1 &  9         &  J2009+7229, 0.77   &  3C286      \\
L1165                                       & A-array & 11 Oct 2012 & 1 &  9         &  J2022+6136, 3.0    &  3C48      \\
CB230                                       & A-array & 02 Nov 2012 & 1 &  9         &  J2009+7229, 0.82   &  3C48      \\

L1165                                       & A-array & 03 Oct 2012 & 1 &  21        &  J2022+6136, 1.96   &  3C48      \\
L1157                                       & A-array & 08 Oct 2012 & 1 &  21        &  J2006+6424, 0.47   &  3C48      \\
CB230                                       & A-array & 09 Oct 2012 & 1 &  21        &  J2009+7229, 0.74   &  3C48      \\
L1165                                       & A-array & 02 Nov 2012 & 1 &  21        &  J2022+6136, 2.01   &  3C48      \\
CB230                                       & A-array & 07 Dec 2012 & 1 &  21        &  J2009+7229, 0.81   &  3C48      \\

L1165                                       & A-array & 13 Oct 2012 & 1 &  41        &  J2022+6136, 0.86   &  3C48      \\
L1165                                       & A-array & 15 Oct 2012 & 1 &  41        &  J2022+6136, 0.91   &  3C48      \\
L1165                                       & A-array & 15 Oct 2012 & 1 &  41        &  J2022+6136, 0.91   &  3C48      \\
L1157                                       & A-array & 03 Nov 2012 & 1 &  41        &  J2006+6424, 0.51   &  3C48      \\
L1157                                       & A-array & 09 Nov 2012 & 1 &  41        &  J2006+6424, 0.44   &  3C48      \\
L1157                                       & A-array & 14 Nov 2012 & 1 &  41        &  J2006+6424, 0.42   &  3C48      \\
L1157                                       & A-array & 20 Nov 2012 & 1 &  41        &  J2006+6424, 0.42   &  3C48      \\
L1157                                       & A-array & 20 Nov 2012 & 1 &  41        &  J2006+6424, 0.42   &  3C48      \\
CB230                                       & A-array & 28 Nov 2012 & 1 &  41        &  J2009+7229, 0.69   &  3C48      \\
CB230                                       & A-array & 30 Nov 2012 & 1 &  41        &  J2009+7229, 0.7    &  3C286      \\
CB230                                       & A-array & 02 Dec 2012 & 1 &  41        &  J2009+7229, 0.7    &  3C48      \\
CB230                                       & A-array & 05 Dec 2012 & 1 &  41        &  J2009+7229, 0.74   &  3C48      \\
CB230                                       & A-array & 05 Dec 2012 & 1 &  41        &  J2009+7229, 0.75   &  3C48      \\
CB230, L1165                                & C-array & 07 Jul 2013 & 1.75 &  4.5, 7.4 & J2148+6107, 1.2   &  3C48      \\
\enddata
\tablecomments{We observed in fast-switching mode for the B and A configuration data and due to non-optimal calibrator separations, our on-source
efficiency was generally ~20-25\%; our efficiency was $\sim$50\% in C configuration. The B and C configuration data were from the
VLA project 12A-082 and the A configuration data were from 12B-211.}
\tablenotetext{*}{Scans for the flux calibrator were missing from this observation and the flux density from the 19 Feb track was assumed.}
\tablenotetext{\dagger}{Incorrect coordinates were used for CB230 IRS1 and the source was at the edge of the primary beam,
 these data are not used in analysis.}
\tablenotetext{a}{This track had very high phase variance; however, we were able to use the calibrator data to determine the flux density
for application to the 12 Feb track.}
\tablenotetext{b}{Problems with pointing caused the first 2 hours of the data to be unusable.}

\end{deluxetable}

\begin{deluxetable}{llllllll}
\tablewidth{0pt}
\rotate
\tabletypesize{\scriptsize}
\tablecaption{CARMA Observation Log}
\tablehead{
  \colhead{Sources} &\colhead{Config.}  & \colhead{Date} & \colhead{Track Length} & \colhead{Central Frequency(s)} & \colhead{$\tau_{225GHz}$} & \colhead{Gain Calibrator/Flux Density} & \colhead{Flux Calibrator}\\
                    &                   & \colhead{(UT)} &  \colhead{(hrs)}       & \colhead{(GHz)}                  & \colhead{(nepers)}          &  \colhead{(Name, Jy)}                  &\\
}
\startdata
L1165 & C-array & 9 Dec 2012 & 4   &  225.0491 & 0.3 &  3C418, 1.0   &  Neptune      \\
L1165 & B-array & 18 Jan 2013 & 8 &  225.0491 & 0.07 &  3C418, 1.0   &  MWC349     \\
L1165 & B-array & 19 Jan 2013 & 4.25 &  225.0491 & 0.06 &  3C418, 1.0  &  Neptune     \\
\\
CB230 & C-array & 8 Jan 2013 & 4   &  90.9027 & 0.35 &  1927+739, 3.7   &  MWC349      \\
CB230 & C-array & 27 May 2013 & 4.25   &  90.9027 & 0.25 &  1927+739, 4.4   &  MWC349      \\
CB230 & C-array & 28 May 2013 & 4.5   &  90.9027 & 0.4 &  1927+739, 4.1   &  Neptune      \\
CB230 & C-array & 29 May 2013 & 6.0   &  90.9027 & 0.65 &  1927+739, 4.2   &  MWC349      \\
\\

\enddata
\end{deluxetable}

\begin{deluxetable}{lllllllllllll}
\tablewidth{0pt}
\rotate
\tabletypesize{\tiny}
\tablecaption{L1165-SMM1 Measurements}
\tablehead{
  \colhead{Wavelength} & \colhead{Configuration(s)} & \colhead{Integrated Flux} & \colhead{Peak Flux} & \colhead{Aperture} & \colhead{Gaussian Size} & \colhead{Gaussian PA} & \colhead{Deconvolved Size} & \colhead{Deconvolved PA} & \colhead{Robust} & \colhead{Taper} & \colhead{Beam} & \colhead{Beam PA}\\
  \colhead{(mm)}        &                           & \colhead{(mJy)}           & \colhead{(mJy)}    & \colhead{(\arcsec)}& \colhead{(\arcsec)}    & \colhead{(\degr)}& \colhead{(\arcsec)}    & \colhead{(\degr)}  & &\colhead{(k$\lambda$)}  &\colhead{(\arcsec)} & \colhead{(\degr)} \\
}
\startdata
Total \\
1.3 & BC  & 82.3$\pm$3   & 52.8$\pm$1   & 0.9 $\times$ 0.9   & 0.551 $\times$ 0.331 & -55.1 & 0.40 $\times$ 0.12  & -48.4 & 1 & none  & 0.39 $\times$ 0.30   & -72.0\\
1.3 & B   & 69.9$\pm$11    & 44.5$\pm$3   & 0.9 $\times$ 0.9   & 0.443 $\times$ 0.266 & -55.5 & 0.35 $\times$ 0.12  & -0.1 & -2 & none  & 0.28 $\times$ 0.23   & -80.9\\
3.4 & CD  & 13.3$\pm$0.8   & 10.3$\pm$0.3   & 10 $\times$ 10     & 3.89 $\times$ 2.93   & -85.6 & 1.81 $\times$ 0.990 & 85.6  & 2 & none  & 3.45 $\times$ 2.74   & -81.0\\
7.3 & A   & 0.9$\pm$0.1   & 0.34$\pm$0.03 & 0.4 $\times$ 0.25  &  ...                  & ...   & ...                 & ...   & 2 & none  & 0.072 $\times$ 0.057 & -32.2\\
7.3 & ABC & 1.4$\pm$0.1   & 0.45$\pm$0.02 & 0.4 $\times$ 0.25  & ...                  & ...   & ...                 & ...   & 2 & none  & 0.089 $\times$ 0.074 & -30.2\\
7.3 & ABC & 1.7$\pm$0.2   & 0.51$\pm$0.02 & 0.4 $\times$ 0.25  & ...                  & ...   & ...                 & ...   & 2 & 2000  & 0.13 $\times$ 0.11   & -12.0\\
7.3 & ABC & 1.4$\pm$0.1   & 0.71$\pm$0.03 & 0.9 $\times$ 0.9   & 0.46 $\times$ 0.31   & 135.9 & 0.35 $\times$ 0.12  & 134.3 & 2 & 500   & 0.30 $\times$ 0.29   & -22.2\\
14  & AB  & 0.35$\pm$0.02    & 0.21$\pm$0.01  & 0.9 $\times$ 0.9   &...                 & ...   & ...                 & ...   & 2 & 1000  & 0.19 $\times$ 0.16   & 13.7\\
33  & AB  & 0.25$\pm$0.01 & 0.09$\pm$0.01 & 0.9 $\times$ 0.9  &...                & ...   & ...                 & ...   & 0.5 & none & 0.28 $\times$ 0.20  & 30.3\\
40 & C   & 0.17$\pm$0.02  & 0.14$\pm$0.01  & 6 $\times$ 6      &...    & ... & ...  & ...   & 1 & none & 3.7 $\times$ 2.7  & 1.2\\
65  & C  & 0.13$\pm$0.03  & 0.11$\pm$0.01  & 10 $\times$ 10    & ...   & ... & ...  & ...   & 1 & none & 6.0 $\times$ 4.5  & -178.6\\

Primary\\
1.3 & BC  & 59.8$\pm$3    & 44.6$\pm$2   & Gaussian fit      & 0.342 $\times$0.245  & 135.0$\pm$3.5  & ...   & ...       & -2 & none    & 0.28 $\times$ 0.23  & -81.0\\
7.3 & A   & 0.45$\pm$0.2    & 0.34$\pm$0.03 & 0.25 $\times$ 0.25  & 0.13 $\times$ 0.063  & 120.0 & 0.11 $\times$ 0.018 & 116.0 & 2 & none  & 0.072 $\times$ 0.057 & -32.2\\
7.3 & ABC & 1.0$\pm$0.05    & 0.45$\pm$0.02 & 0.25 $\times$ 0.25  & 0.15 $\times$ 0.096  & 125.5 & 0.12 $\times$ 0.056 & 120.7 & 2 & none  & 0.089 $\times$ 0.074 & -30.2\\
7.3 & ABC & 0.76$\pm$0.05   & 0.51$\pm$0.02 & 0.25 $\times$ 0.25  & 0.18 $\times$ 0.13   & 140.0 & 0.13 $\times$ 0.048 & 130.1 & 2 & 2000  & 0.13 $\times$ 0.11   & -12.0\\
14  & AB  & 0.23$\pm$0.03  & 0.18$\pm$0.01 & 0.25 $\times$ 0.25  & 0.21 $\times$ 0.19   & 156.2 & 0.12 $\times$ 0.050  & 130  & 2 & 1000  & 0.19 $\times$ 0.16   & 13.7\\
33  & AB  & 0.10$\pm$0.01  & 0.09$\pm$0.01 & 0.3 $\times$ 0.3  & 0.33 $\times$ 0.22    & 41.5  & 0.19 $\times$ 0.070   & 56  & 0.5 & none & 0.28 $\times$ 0.20  & 30.3\\

Secondary\\
1.3 & BC  & 11.8$\pm$3   & 11.2$\pm$2     & Gaussian fit       & 0.298 $\times$ 0.221 & 173$\pm$15  & ...                 & ...  & -2   & none  & 0.28 $\times$ 0.23  & -81.0\\
7.3 & A   & 0.17$\pm$0.06   & 0.12$\pm$0.03   & 0.25 $\times$ 0.25 & ...                  &...          & ...                 & ...  & 2    & none  & 0.072 $\times$ 0.057 & -32.2\\
7.3 & ABC & 0.34$\pm$0.04  & 0.17$\pm$0.02    & 0.25 $\times$ 0.25 & 0.17 $\times$ 0.14   & 127.5       & 0.15 $\times$ 0.11  & 123  & 2    & none  & 0.089 $\times$ 0.074 & -30.2\\
7.3 & ABC & 0.34$\pm$0.05   & 0.19$\pm$0.02   & 0.25 $\times$ 0.25 & 0.22 $\times$ 0.184  & 151.1       & 0.17 $\times$ 0.14  & 141   & 2   & 2000  & 0.13 $\times$ 0.11  & -12.0\\
14  & AB  & 0.01$\pm$0.02 & 0.069$\pm$0.01  & 0.25 $\times$ 0.25 & 0.28 $\times$ 0.21   & 40.1        & 0.21 $\times$ 0.12  & 49.1  & 2   & 1000  & 0.19 $\times$ 0.16  & 13.7\\
33  & AB  & 0.03$\pm$0.02 & 0.036$\pm$0.01  & 0.25 $\times$ 0.25 & 0.32 $\times$ 0.235  & 38.5        & ...                 & ...   & 0.5 & none  & 0.28 $\times$ 0.20  & 30.3\\

\enddata
\tablecomments{Uncertainties on the integrated and peak flux densities are from the flux measurements in the specified apertures, unless `Gaussian fit' is
listed in the Aperture column. We do not give uncertainties on the Gaussian fit parameters for the sake of brevity.}
\end{deluxetable}

\begin{deluxetable}{lllllllllllll}
\tablewidth{0pt}
\rotate
\tabletypesize{\tiny}
\tablecaption{CB230 IRS1 Measurements}
\tablehead{
  \colhead{Wavelength} & \colhead{Configuration(s)} & \colhead{Integrated Flux}   & \colhead{Peak Flux} & \colhead{Aperture} & \colhead{Gaussian Size} & \colhead{Gaussian PA} & \colhead{Deconvolved Size} & \colhead{Deconvolved PA} & \colhead{Robust} & \colhead{Taper} & \colhead{Beam} & \colhead{Beam PA}\\
  \colhead{(mm)}        &                           & \colhead{(mJy)}           & \colhead{(mJy)}    & \colhead{(\arcsec)} & \colhead{(\arcsec)}    & \colhead{(\degr)}& \colhead{(\arcsec)}    & \colhead{(\degr)}  & &\colhead{(k$\lambda$)}  &\colhead{(\arcsec)} & \colhead{(\degr)} \\
}
\startdata
CB230 IRS1 \\

3.4 & C  & 13.3$\pm$0.8    & 10.3$\pm$0.3   & 10 $\times$ 10      & 3.89 $\times$ 2.93   & -85.6 & 1.81 $\times$ 0.990 & 85.6  & 2 & none  & 3.45 $\times$ 2.74   & -81.0\\
7.3 & A   & 1.5$\pm$0.2   & 0.42$\pm$0.02   & 0.4 $\times$ 0.25 & ...                  & ...   & ...                 & ...   & 2 & none  & 0.068 $\times$ 0.051 & -23.9\\
7.3 & ABC & 1.8$\pm$0.1   & 0.46$\pm$0.02   & 0.4 $\times$ 0.25 & ...                  & ...   & ...                 & ...   & 2 & none  & 0.073 $\times$ 0.055 & -21.0\\
7.3 & ABC & 1.5$\pm$0.1   & 0.58$\pm$0.03   & 0.4 $\times$ 0.25 & ...                  & ...   & ...                 & ...   & 2 & 2000  & 0.097 $\times$ 0.086   & -6.1\\
7.3 & ABC & 1.5$\pm$0.1   & 0.79$\pm$0.04   & 0.9 $\times$ 0.9  & 0.55 $\times$ 0.29   & 54.8 & 0.48 $\times$ 0.14  & 56.6 & 2 & 500   & 0.27 $\times$ 0.25   & 17.88\\
14  & AB  & 0.34$\pm$0.03  & 0.17$\pm$0.01    & 0.9 $\times$ 0.9 & ...                 & ...   & ...                 & ...   & 2 & 1000  & 0.18 $\times$ 0.17   & -12.7\\
33  & AB  & 0.1$\pm$0.02  & 0.053$\pm$0.01  & 0.5 $\times$ 0.5 & ...                & ...   & ...                 & ...   & 1 & none & 0.27 $\times$ 0.24  & 16.8\\
40 & C   & 0.098$\pm$0.02 & 0.071$\pm$0.01  & 6 $\times$ 6   & ...    & ... & ...  & ...   & 1 & none & 4.1 $\times$ 2.7  & -16.2\\
65  & C  & 0.064$\pm$0.02 & 0.064$\pm$0.01  & 8 $\times$ 8   &  ...   & ... & ...  & ...   & 1 & none & 6.6 $\times$ 4.5  & -15.8\\

Primary\\

7.3 & A   & 0.94$\pm$0.1	   & 0.42$\pm$0.02   & 0.25 $\times$ 0.25  & 0.09 $\times$ 0.076  & 3.4    & 0.064 $\times$ 0.047 & 335.0 & 2 & none  & 0.068 $\times$ 0.051 & -23.9\\
7.3 & ABC & 1.1$\pm$0.1     & 0.46$\pm$0.02   & 0.25 $\times$ 0.25  & 0.10 $\times$ 0.086  & 179.3  & 0.074 $\times$ 0.061 & 31 & 2 & none & 0.073 $\times$ 0.055 & -21.0\\
7.3 & ABC & 0.95$\pm$0.1    & 0.58$\pm$0.03    & 0.25 $\times$ 0.25 & 0.12 $\times$ 0.11   & 180.0  & 0.075 $\times$ 0.07  & 19 & 2 & 2000  & 0.097 $\times$ 0.086   & -6.1\\
14  & AB  & 0.26$\pm$0.03   & 0.18$\pm$0.01   & 0.25 $\times$ 0.25  & 0.24 $\times$ 0.20   & 172.9  &  0.16 $\times$ 0.10  & 173.6  & 2 & 1000  & 0.18 $\times$ 0.17   & -12.7\\
33  & AB  & 0.054$\pm$0.003 & 0.053$\pm$0.005  & 0.25 $\times$ 0.25 & 0.28 $\times$ 0.25   & 165.5  & ...                  & ...  & 1 & none & 0.27 $\times$ 0.24  & 16.8\\

Secondary\\

7.3 & A   & 0.46$\pm$0.1   & 0.20$\pm$0.02   & 0.25 $\times$ 0.25  & 0.13 $\times$ 0.07  & 67.0  & 0.12 $\times$ 0.02 & 67     & 2   & none  & 0.068 $\times$ 0.051 & -23.9\\
7.3 & ABC & 0.63$\pm$0.1   & 0.22$\pm$0.02   & 0.25 $\times$ 0.25  & 0.13 $\times$ 0.08  & 68.4  & 0.12 $\times$ 0.04 & 68.5   & 2 & none  & 0.073 $\times$ 0.055 & -21.0\\
7.3 & ABC & 0.52$\pm$0.1   & 0.27$\pm$0.03   & 0.25 $\times$ 0.25  & 0.14 $\times$ 0.10  & 66.0  & 0.12 $\times$ 0.03  & 68.7   & 2  & 2000  & 0.097 $\times$ 0.086   & -6.1\\
14  & AB  & 0.092$\pm$0.02 & 0.074$\pm$0.01  & 0.25 $\times$ 0.25  & 0.36 $\times$ 0.19  & 61.2  & 0.31 $\times$ 0.11  & 61.6  & 2  & 1000  & 0.18 $\times$ 0.17   & -12.7\\
33  & AB  & 0.0093$\pm$0.008 & 0.018$\pm$0.005 & 0.25 $\times$ 0.25  & ...  & ... & ... & ...  & 1.0 & none & 0.27 $\times$ 0.24  & 16.8\\

CB230 IRS2 \\

3.4 & C & $<$1.0\\
7.3 & ABC & $<$0.08\\
14 & AB & $<$0.04\\
33 & AB & $<$0.02\\
40& AB & $<$0.02\\
65 & AB & $<$0.04\\
\enddata
\tablecomments{Uncertainties on the integrated and peak flux densities are from the flux measurements in the specified apertures, unless `Gaussian fit' is
listed in the Aperture column. We do not give uncertainties on the Gaussian fit parameters for the sake of brevity.}

\end{deluxetable}

\begin{deluxetable}{lllllllllllll}
\tablewidth{0pt}
\rotate
\tabletypesize{\tiny}
\tablecaption{L1157-mm Measurements}
\tablehead{
  \colhead{Wavelength} & \colhead{Configuration(s)} & \colhead{Integrated Flux} & \colhead{Peak Flux} & \colhead{Aperture} & \colhead{Gaussian Size} & \colhead{Gaussian PA} & \colhead{Deconvolved Size} & \colhead{Deconvolved PA} & \colhead{Robust} & \colhead{Taper} & \colhead{Beam} & \colhead{Beam PA}\\
  \colhead{(mm)}        &                           & \colhead{(mJy)}           & \colhead{(mJy)}    & \colhead{(\arcsec)}& \colhead{(\arcsec)}    & \colhead{(\degr)}& \colhead{(\arcsec)}    & \colhead{(\degr)}  & &\colhead{(k$\lambda$)}  &\colhead{(\arcsec)} & \colhead{(\degr)} \\
}
\startdata
Total \\
1.3 & B  & 181$\pm$30      & 98.5$\pm$1.0   & 1 $\times$ 1      & 0.509 $\times$ 0.37   & 80.8 & 0.35 $\times$ 0.20 & 70.6  & 2 & none  & 0.37 $\times$ 0.33   & -83.2\\
3.4 & A  & 9.0$\pm$1.0     & 6.8$\pm$0.3   & 1 $\times$ 1      & 0.48 $\times$ 0.40   & 75.0 & 0.33 $\times$ 0.171 & 59.9  & 2 & none  & 0.39 $\times$ 0.33   & -61.4\\
7.3 & A   & 1.7$\pm$0.1   & 0.49$\pm$0.02  & 0.3 $\times$ 0.3   & 0.11 $\times$ 0.11 & 130.0   & 0.093 $\times$ 0.083  & 80  & 2 & none  & 0.070 $\times$ 0.052 & -23.2\\
7.3 & ABC & 2.2$\pm$0.1   & 0.71$\pm$0.02  & 0.3 $\times$ 0.3   & 0.12 $\times$ 0.12   & 78.0   & 0.11 $\times$ 0.082    & 72.0   & 2 & none  & 0.073 $\times$ 0.055 & -21.0\\
7.3 & ABC & 2.2$\pm$0.1   & 1.0$\pm$0.02   & 0.3 $\times$ 0.3   & 0.15 $\times$ 0.15  & 8.4   & 0.11 $\times$ 0.10  & 56   & 2 & 2000  & 0.11 $\times$ 0.095   & -13.2\\
7.3 & ABC & 2.4$\pm$0.1   & 1.7$\pm$0.03  & 0.5 $\times$ 0.5   & 0.35 $\times$ 0.33  & 11.3   & 0.20 $\times$ 0.18  & 31   & 2 & 500  & 0.29 $\times$ 0.28   & 0.7\\
14  & AB  & 0.54$\pm$0.07 & 0.31$\pm$0.01  & 0.5 $\times$ 0.5   & 0.21 $\times$ 0.14  & 134.1 & 0.12 $\times$ 0.069  & 134.3   & 2 & 1000  & 0.18 $\times$ 0.17   & -12.7\\
33  & AB  & 0.24$\pm$0.02 & 0.21$\pm$0.01 & 0.5 $\times$ 0.5   & 0.36 $\times$ 0.25              & 35.33   & ...               & ...   & 2 & none & 0.36 $\times$ 0.23  & 40.6\\
36 & C   & 0.21$\pm$0.07       & 0.21$\pm$0.02  &  10 $\times$ 10      &3.3 $\times$ 2.5    & 164.6 & ...  & ...   & 2 & none & 3.5 $\times$ 2.5  & -37.3\\
62  & C  & 0.18$\pm$0.05       & 0.19$\pm$0.03  &  10 $\times$ 10      & 5.7 $\times$ 4.8    & 130.7 & ...  & ...   & 2 & none & 6.6 $\times$ 4.6  & -54.9\\

\enddata
\tablecomments{Uncertainties on the integrated and peak flux densities are from the flux measurements in the specified apertures, unless `Gaussian fit' is
listed in the Aperture column. We do not give uncertainties on the Gaussian fit parameters for the sake of brevity.}

\end{deluxetable}

\begin{deluxetable}{lllll}
\tablewidth{0pt}
\tabletypesize{\scriptsize}
\tablecaption{Spectral Slopes}
\tablehead{
  \colhead{Source} & \colhead{Free-free Slope} & \colhead{F$_0$(free-free)} & \colhead{Thermal Slope} & \colhead{F$_0$(thermal)}\\
  \colhead{}        &                           & \colhead{(mJy)}           &                         & \colhead{(mJy)}  
}
\startdata
CB230             &           -0.71$\pm$0.42   &  1.2$\pm$2.0 &      -3.3$\pm$0.38   & 890$\pm$450        \\
L1157             &           -0.35$\pm$0.42   &  0.77$\pm$1.2 &      -2.6$\pm$0.25   & 270$\pm$80         \\
L1165             &           -0.013$\pm$0.33   &  0.15$\pm$0.26   &      -2.45$\pm$0.13   & 160$\pm$20        \\
\enddata
\tablecomments{The spectral slopes are defined by the convention F$_{\lambda}$ = F$_{0}(\lambda/ 1 mm)^{\alpha}$ where
$\alpha$ is the spectral slope. The thermal and free-free spectral indicies are derived from simultaneous fitting of
the two components. }

\end{deluxetable}

\begin{deluxetable}{llllc}
\tablewidth{0pt}
\tabletypesize{\scriptsize}
\tablecaption{Source Masses from Dust Emission}
\tablehead{
  \colhead{Source} & \colhead{Wavelength} & \colhead{$\beta$} & \colhead{Mass}  & Fraction of \\
  \colhead{}        &\colhead{(mm)}       &                   & \colhead{($M_{\sun}$)} & Free-Free Emission   
}  

\startdata
CB230 IRS1 A+B&     3.4 &  1.3 $\pm$ 0.38   & 0.13 $\pm$ 0.008   & 0.038  \\
CB230 IRS1 A+B&     7.3 &  " "             & 0.16  $\pm$  0.01    & 0.19  \\
CB230 IRS1 A+B&     14 &  " "              & 0.18  $\pm$  0.02     & 0.53 \\
CB230 IRS1 A  &     7.3 &  " "             & 0.054 $\pm$ 0.004   & 0.18  \\
CB230 IRS1 A  &      14 &  " "             & 0.075 $\pm$ 0.008   & 0.40  \\
CB230 IRS1 B  &     7.3 &  " "             & 0.034 $\pm$ 0.005   & 0.052    \\
CB230 IRS1 B  &     14 &  " "              & 0.036 $\pm$ 0.008   & 0.19    \\
\\
L1165-SMM1 A+B&     1.3 &  0.45 $\pm$ 0.13  & 0.027 $\pm$ 0.001   & 0.002   \\
L1165-SMM1 A+B&     7.3 &  " "            & 0.027 $\pm$ 0.002   & 0.10  \\
L1165-SMM1 A+B&     14 &  " "             & 0.022 $\pm$ 0.003   & 0.41  \\
L1165-SMM1 A  &     1.3 &  " "            & 0.020 $\pm$ 0.001   & 0.002   \\
L1165-SMM1 A  &     7.3 &  " "            & 0.014 $\pm$ 0.0009  & 0.15    \\
L1165-SMM1 A  &     14 &  " "             & 0.012 $\pm$ 0.0015  & 0.50   \\
L1165-SMM1 B  &     1.3 &  " "            & 0.0038 $\pm$ 0.0008 & 0.003  \\
L1165-SMM1 B  &     7.3 &  " "            & 0.0065 $\pm$ 0.001  & 0.10   \\
L1165-SMM1 B  &     14 &  " "             & 0.0067 $\pm$ 0.002  & 0.33   \\
\\
L1157-mm      &     1.3 &  0.6 $\pm$ 0.25   & 0.063 $\pm$ 0.01    & 0.004    \\
L1157-mm      &     3.4 &  " "            & 0.034 $\pm$ 0.004   & 0.06  \\
L1157-mm      &     7.3 &  " "            & 0.058 $\pm$ 0.002   & 0.16 \\
L1157-mm      &     14 &  " "             & 0.026 $\pm$ 0.004   & 0.65 \\
\enddata
\tablecomments{The calculated masses all assume a dust-to-gas mass ratio of 1:100 and a dust 
temperature of 30 K. The flux densities used for the total masses at 7.3 mm are derived from the 
images tapered to 500 k$\lambda$. The uncertainties on the masses are statistical only; the 1.3 mm and 3.4 mm
measurements carry an absolute calibration accuracy of 10\% - 20\% and the 7.3 mm data have
a calibration accuracy of $\sim$10\%. Note that there are additional systematic uncertainties
from the fits to the spectral slopes that have not been accounted for. }

\end{deluxetable}

\end{document}